\documentclass[aps,prb,twocolumn,showpacs]{revtex4}

\usepackage{graphicx,amsmath,color}

\newcommand{\eq}[1]{Eq.~(\ref{#1})}
\newcommand{\fig}[1]{Fig.~\ref{#1}}

\newcommand{\olcite}[1]{Ref.~\onlinecite{#1}}
\newcommand{\olcites}[1]{Refs.~\onlinecite{#1}}

\newcommand{\eeq}{ \end{equation} }
\newcommand{\beq}{ \begin{equation} }

\newcommand{\eea}{ \end{eqnarray} }
\newcommand{\bea}{ \begin{eqnarray} }

\newcommand{\etal}{ {\em et al.}}
\newcommand{\bhua}{ {\bf \hat{u}}_{1} }
\newcommand{\bhub}{ {\bf \hat{u}}_{2} }
\newcommand{\bhui}{ {\bf \hat{u}}_{i} }
\newcommand{\bhu}{ {\bf \hat{u}} }
\newcommand{\bu}{ {\bf u} }
\newcommand{\bra}{ {\bf r}_1 }
\newcommand{\brb}{ {\bf r}_2 }
\newcommand{\br}{ {\bf r} }
\newcommand{\bz}{ {\bf \hat{z}} }
\newcommand{\bn}{ {\bf \hat{n}} }
\newcommand{\bx}{ {\bf \hat{x}} }
\newcommand{\bv}{ {\bf \hat{v}} }
\newcommand{\bw}{ {\bf \hat{w}} }
\newcommand{\kbt}{k_{\rm B}T}
\newcommand{\nab}{  {\bf \nabla }}

\begin{document}

\title{Generalized van der Waals theory for the twist elastic modulus and helical pitch of cholesterics}
\author{H. H. Wensink }
\email{r.wensink@imperial.ac.uk}
\author{G. Jackson}
\affiliation{Department of Chemical Engineering, Imperial College London, South Kensington Campus, London SW7 2AZ, United Kingdom}

\date{\today}

\begin{abstract}

We present a generalized van der Waals theory for a lyotropic cholesteric 
system of chiral spherocylinders based on the classical Onsager theory for 
hard anisometric bodies. The rods consist of a hard spherocylindrical 
backbone surrounded with a square-well potential to account for attractive 
(or soft repulsive) interactions. Long-ranged chiral interactions are 
described by means of a simple pseudo-scalar potential which is 
appropriate for weak chiral forces of a predominant electrostatic origin. 
Based on the formalism proposed by Straley [Phys. Rev. A {\bf 14}, 1835 
(1976)] we derive explicit algebraic expressions for the twist elastic 
modulus and the cholesteric pitch for rods as a function of density and 
temperature. The pitch varies non-monotonically with density, with a sharp 
decrease at low packing fractions and a marked increase at higher packing 
fractions. A similar trend is found for the temperature dependence. The 
unwinding of the helical pitch at high densities (or low temperatures) 
originates from a marked increase in the local nematic order and a steep 
increase of the twist elastic resistance associated with near-parallel 
local rod configurations. This contrasts with the commonly held view that 
the increase in pitch with decreasing temperature as often observed in 
cholesterics is due to layer formation resulting from pre-smectic 
fluctuations.  The increase in pitch with increasing temperature is 
consistent with an entropic unwinding as the chiral interaction becomes 
less and less significant than the thermal energy. The variation of the 
pitch with density, temperature and contour length is in qualitative 
agreement with recent experimental results on colloidal {\em fd} rods.

\end{abstract}


\maketitle

\section{Introduction}

Cholesteric liquid crystals (LCs) display a variety of extraordinary features due to the
existence of a mesoscopic helical structure which is best known for its
exceptionally large optical rotational power employed in LC display
technology. In contrast to a common nematic phase, where the nematic
director is homogeneous throughout the system, the cholesteric (chiral nematic) phase is
characterized by a helical arrangement of the director field along a
common pitch axis. As a result, the cholesteric phase possesses an
additional mesoscopic length scale, commonly referred to as the `pitch length', which
characterizes the distance along the pitch axis over which the local director
makes a full revolution \cite{gennes-prost}.

Derivatives of cholesterol, chiral molecules which were the first mesogenic substances to be recognized in studies of the melting
point and optical properties of carrot extracts by Reinitzer \cite{reinitzer1888} and Lehmann \cite{lehmann1889},
 belong to the thermotropic class of liquid crystals where phase
transitions are brought about by a variation in the temperature; a large number of thermotropic mesogens with a low to moderate molecular weight that form chiral nematic phases have now been isolated or synthesized.  Their widespread use in optoelectronic applications (e.g., twisted nematic liquid crystal displays in laptop computers, televisions, and mobile phones) is a direct consequence of the unique rheological, electrical and optical properties imparted by the chiral structures.

Lyotropic chiral systems, involving high molecular-weight particles in
solution where the ordering behavior is primarily governed by the
solute concentration, are also common.  Examples are bio-colloidal systems such as DNA
\cite{robinson-pblg,livolantDNAoverview} and the rod-like $fd$-virus \cite{dogic-fraden_fil},
stiff polymers such as polypeptides \cite{uematsu,dupresamulski}, polysaccharides
\cite{sato-teramoto} or cellulose derivatives \cite{gray-cellulose}, and
chiral micelles \cite{hiltrop}. In these systems, the
cholesteric pitch is very sensitive to  concentration, temperature as
well as the solvent conditions such as ionic strength and pH. The dependence of the
pitch upon these
variables has been the subject of intense experimental research
(see for example references \cite{rill-dna,yu-dna,strey-dna, dogic-fraden_chol,grelet-fraden_chol,robinson-pblg, dupreduke75,yoshiba-sato,dong-gray-cellulose,miller-cellulose,microcellulose,sato-sato}).

Theoretical attempts to predict the behavior of the cholesteric pitch are
challenging owing to the complexity of the underlying chiral interaction \cite{harris-rmp} and the inhomogeneous and anisotropic nature of the phase.
Course-grained model potentials aimed
at capturing the essentials of the complex molecular nature of the electrostatics of the surface of such macromolecules have been devised mainly for DNA  \cite{parsegian-podgornik,kornyshevleikin-chiral-err,kornyshevleikin-prl, kornyshevleikin-pitch,tombolatoferrarini}. A more general electrostatic model potential for chiral interactions was proposed much earlier by Goossens \cite{goossens} based on
a spatial arrangement of dipole-dipole and dipole-quadrupole interactions
which can be cast into a multipole expansion in terms of tractable pseudo-scalar
potentials \cite{meervertogen}. This type of electrostatic description of the chiral interaction can be combined with a Maier-Saupe mean-field treatment (see, for example references~\cite{meervertogenJCP, lin-liu77,hu,osipov,kapanowski,emelyanenko}, or with a bare hard-core model and treated within the seminal theory of Onsager~\cite{onsager} (as in the recent study of chiral hard spherocylinders for systems with perfect local nematic order \cite{vargachiral1}).

Taking the alternative view of a steric origin for the chiral interactions, Straley \cite{straleychiral} combined
a leading order pseudo-scalar form with a hard rod model to create a
basic chiral model (hard threaded rod) for lyotropic cholesterics. Though influenced by the work of Goossens on electrostatic forces, the potential used by Straley is
appropriate for short-ranged steric chiral interactions mediated
by a thin helical thread enveloping each rod.  By extending the theory of
Onsager \cite{onsager} to systems with non-uniform director
fields,  microscopic expressions for the macroscopic twist energy
associated with a bulk cholesteric structure can be
deduced. The theoretical treatment was later elaborated in a paper by Odijk
\cite{odijkchiral} leading to  microscopic scaling expressions
for the twist elastic constant and helical pitch of both rigid and
semi-flexible rods. Similar relations, albeit with different scaling
exponents, were obtained by Pelcovits \cite{pelcovits} based on a
corkscrew model.

At this stage one should also acknowledge the related studies on the link between the orientational order parameters and the elastic constants in nematics, which have been influential in shaping some of the theories of chiral phases.  A mean-field treatment of particles with electrostatic chirality but no shape anisotropy has been developed~\cite{saupe,nehring,vandermeer}, as have Onsager DFT theories for anisotropic hard rod-like particles \cite{priest,straley,stecki1,lee-elastic} and hard rods with attractive mean-fields 
\cite{stecki2,gelbart}, the latter being closely related to the achiral contribution of the model employed in our study. The principal finding of this body of work is that the elastic constants are predicted to be proportional to the square of the nematic order parameter, at least for system with weak orientational order, confirming experimental observation \cite{vertogendejeu} and the results of molecular simulation \cite{steuer}. Such a treatment has also been extended to a description of the elastic constants in smectic phases \cite{vdmeerpostma,govers}.

The underlying microscopic physical feature (steric or electrostatic) responsible for the formation of chiral nematic phases is still a matter of debate and controversy. In has been known for some time that the nature of the solvent can have a dramatic effect on the pitch of  cholesteric phases, and can even reverse the sense of the twist (e.g., see the study of Robinson \cite{robinson-pblg}  on poly-$\gamma$-L-benzyl-L-glutamate PBLG, a synthetic polypeptide with an $\alpha$ helical conformation, in achiral solvents such as dioxane and dichloromethane), hinting to a solvent effect of electrostatic origin. In a beautifully revealing but rather overlooked paper, Coates and Gray \cite{coates} showed that the replacement of a hydrogen by a deuterium atom on the carbon backbone of an originally achiral thermotropic mesogen is sufficient to induce a cholesteric structure; the carbon-hydrogen and carbon-deuterium bond lengths are both 1.085 $\text{\AA}$  so one would expect the ``steric shape'' of both molecules to be very similar, indicating that in this case at least the chirality is of a weak and subtle electrostatic nature.

 The fact that the pitch of the cholesteric phase in aqueous solutions of the filamentous $fd$ virus is very sensitive to the ionic strength of the medium but still persists after coating the virus with polyethylene oxide polymer (which would mask any short-range chirality in the particle shape) is also indicative of an electrostatic origin to the interparticle chiral interaction in such lyotropic systems \cite{grelet-fraden_chol,tomar}. This lends credence to the use of electrostatic interactions of the type proposed by Goossens \cite{goossens} decorated with an achiral non-spherical core as physically reasonable microscopic models for chiral systems; an additional repulsive steric chiral core could of course also be incorporated in a more realistic treatment (e.g., see reference \cite{tombolato-grelet}), but this is beyond the scope of our work. 


The helical pitch of cholesteric thermotropic mesogens is almost invariably found to be a
decreasing function of temperature~\cite{gennes-prost}, which is commonly attributed to the presence of an underlying smectic-A phase at lower temperatures: a twisted chiral structure is incommensurate with the smectic layering leading to an unwinding of the pitch as one approaches the transition~\cite{alben-pitch,pindak}. What is surprising is that this thermally induced decrease in the pitch can occur over many decades in temperature, where one would not expect effects due to pre-transitional smectic order. An increase in pitch with increasing temperature has been reported for the cholesterol ester, cholesteryl
[2-(2-ethoxyethoxy-ethyl] carbonate (CEEC) \cite{harada}; interestingly in the case of CEEC the transition from the isotropic phase to the chiral nematic does not appears to be followed by a transition to a smectic phase with a further decrease in temperature,  but rather to the formation of a crystalline state \cite{durand}. A remarkable sense inversion in the helical pitch with temperature has also been found in thermotropic (solvent free) polypeptides \cite{watanabe} and cellulose derivatives \cite{yamagishi}, and in mixtures of right-handed cholesterol chloride and left-handed cholesterol myristate \cite{sackmann}, indicating a subtle balance in the forces giving rise to chiral phases.

The situation is just as intriguing in the case of lyotropic systems where depending on the range of temperature, both negative and positive slopes of the pitch can be observed. At relatively high temperatures a marked increase of the pitch has been found in polypeptide systems (mixtures of PBLG in dioxane, chloroform, and dichloromethane)
\cite{dupreduke75} and in aqueous solutions of $fd$-virus rods \cite{dogic-fraden_chol}. In the latter system the pitch is found to increase with increasing concentration well before the transition to a smectic phase.
An unwinding of the cholesteric phase (with a corresponding increase in the pitch) is also observed in aqueous solutions of DNA \cite{rill-dna} as the concentration of the macromolecules is increased  towards a high-density hexagonal columnar (positionally ordered) state \cite{DNAcol}; this is analogous to the divergence of the pitch found in thermotropic mesogens as one approaches the smectic state on lowering the temperature.

The early molecular-field approaches based on the Maier-Saupe electrostatic picture of the mesogenic interaction \cite{meervertogenJCP,lin-liu77} fail to give a consistent picture of the variation of the pitch with temperature for cholesteric phases, mainly because of the lack of a hard-core exclude volume contribution in the interaction which plays
an essential role in the stabilization of the orientationally ordered phase \cite{onsager}.

The important role of the mesogen's shape
in understanding the temperature dependence of the pitch  was pointed out early on
by Kimura \etal \cite{kimura-hosino}.
The insensitivity of the pitch to temperature found in lattice simulations of sites interacting through a Goossens electrostatic potential~\cite{saha} supports the view that one requires a balance between the opposing twist elastic forces (which are primarily a consequence of the repulsive interactions) and the chiral torque (due to the cholesteric interactions) to observe the subtle dependencies found for the pitch.
A proper account of the rod-like
backbone turns out to be essential to explain the unwinding of the cholesteric structure with increasing temperature \cite{sato-sato,vargachiral1}.

It is clear from the preceding discussion that a number of important questions remain unresolved in our understanding of the origin of chirality and the related dependence of the helical pitch on the thermodynamic variables such as temperature and composition. Does one require the presence of an underlying smectic (or columnar) phase to observe an increase in pitch with decreasing temperature? Under what conditions would one expect a thermally induced unwinding of the chiral nematic phase and when would a non-monotonic temperature dependence of the pitch be obtained? What are the competing roles of repulsive excluded volume interactions, and isotropic and anisotropic (chiral and achiral) attractive interactions in stabilizing a twisted equilibrium structure? As de Gennes and Prost state in their monograph when referring to the temperature dependence of the pitch: ``Their origin is not yet quite clear''. In this paper, we will build upon the idea of coupling a hard core with chiral electrostatic interactions and present a microscopic
theory for rigid chiral spherocylinders of arbitrary aspect ratio. 

Our
motivation for the analysis is three-fold. First of all, most investigations thus
far have been restricted to infinitely thin rod-like species.
We show that the  rod thickness  plays
 a crucial role in the behavior of the pitch in cholesteric systems
with strong orientational order and helps to account for a
non-monotonic dependence of the pitch with temperature and concentration.
Secondly, the role of long-range chiral and achiral dispersive forces
are investigated here by means of a simple square-well (SW) potential. This allows us to
probe the generic temperature dependence of the pitch for chiral rods
where additional achiral attractions (or soft-repulsions) are at play.  Finally, we pay particular attention to the implications of chirality in the inter-particle interactions on the phase behavior.
Our results are entirely algebraic and contain explicit expressions for the
twist energy, elastic modulus, and cholesteric pitch as a function of density and temperature for a given rod aspect-ratio and set of  intermolecular SW parameters such as the interaction range and the relative magnitude of the chiral and achiral attractive/soft-repulsive rod interactions. The results are fully consistent with experimental results for $fd$-virus rods and may prove helpful in interpreting observations in other cholesteric systems.

This paper is laid out as follows. We start with a brief exposition of
Straley's theory for the deformation free energy of the
cholesteric state in Sec. II. A suitable spherocylinder potential is
introduced in Sec. III and incorporated into the deformation free energy through a van der
Waals treatment based on the Onsager-Parsons theory for anisometric hard
bodies. In Sec. IV, microscopic expressions for the twist
parameters and cholesteric pitch is derived. These are analyzed
and the predictions compared with experimental results in Sec V. Finally, some concluding
remarks are made in Sec. VI. Technical issues will  be
relegated to a number of Appendices.

\section{Microscopic theory for the twisted nematic}

To describe the properties of the cholesteric phase
we will follow closely the analysis proposed by Straley
\cite{straleychiral, allenevans}. The aim is to calculate the distortion
free energy
associated with non-uniform nematic director fields.  In a distorted
(e.g., twisted) nematic phase the director is no longer spatially uniform
but depends on position. The same holds for
the orientational
distribution function  (ODF), $f(\bhu \cdot \bn(\br))$ which describes the  probability
of finding a particle with a given orientational unit vector $\bhu$ with 
respect to a locally varying director $\bn(\br)$. Within an Onsager-type 
formulation \cite{onsager}, the excess Helmholtz free energy of a 
distorted 
nematic state of $N$ particles in a volume $V$ can be cast in the following form:
\beq
F^{\text{ex}}  = \frac{\rho ^2}{2} \int d {\bf s}  \Phi ( \br_{12}  ; 
\bhua, \bhub) f ( \bhua \cdot \bn ( \bra ) ) f ( \bhub \cdot \bn  ( \brb ) )  , \label{fgen}
\eeq
with $\int d {\bf s} = \iiiint d \bra d \brb d \bhua d \bhub $ and  $\rho = N/V$ the number density.  The kernel $\Phi$ accounts for the pair interaction of two rods for a relative centre-of-mass separation $ \br_{12} = \brb - \bra $ and orientations $\bhua$, $\bhub$.
This quantity will be fully specified later. Note that,
ˆstrictly, $f = 1/4\pi $ in the isotropic state where the particle orientations
are completely random and the director field becomes irrelevant.

If the spatial variation of the local nematic director is weak such that 
the associated distortion wavelength is much larger than the particle 
dimensions, the ODF may be approximated by a Taylor expansion. For the 
spatial integration it is expedient to switch to a new coordinate system, 
$\br _{i} \rightarrow {\bf R} \pm \br_{12}/2$ for $i=1,2$, in terms of 
${\bf 
R} = (\bra + \brb)/2$ and the centre-of mass distance $\br_{12}$. 
Expanding the ODF then gives 
\begin{eqnarray} f ( \bhu _{i} \cdot \bn ( 
\br_{i} ) )& = & f( \bhu _{i} \cdot \bn ({\bf R})) \pm \frac{ \br_{12} 
\cdot \nab }{2} f ( \bhu _{i} \cdot \bn ({\bf R}) ) \nonumber \\ && + 
{\cal{O}} \left [ ( \nab \cdot \br_{12} ) ^2 \right ]  . 
\end{eqnarray} 
With this result, the excess free energy for weak director gradients 
becomes after some rearrangement \begin{eqnarray}
 F^{\text{ex}} & = & \frac{\rho ^2}{2}  \int d {\bf s}^{\prime}   \Phi ( 
\br_{12}  ; \bhua, \bhub) f ( \bhua \cdot \bn ({\bf R}))
f ( \bhub \cdot \bn ({\bf R})) \nonumber \\
& + & \frac{\rho ^2}{2}  \int d {\bf s}^{\prime}   \Phi ( \br_{12}  ; \bhua, \bhub)  f ( \bhua \cdot \bn ({\bf R})) \dot{f} ( \bhub \cdot \bn ({\bf R}))     \nonumber \\
&& \times  [ ( \br_{12} \cdot \nab )  \bn ({\bf R}) \cdot \bhub ]
\nonumber \\
& - & \frac{\rho ^2}{4}  \int d {\bf s}^{\prime}   \Phi (  \br_{12}  ; \bhua, \bhub)  \dot{f} ( \bhua \cdot \bn ({\bf R})) \dot{f} ( \bhub \cdot \bn ({\bf R})) \nonumber \\
&&  \times  [ ( \br_{12} \cdot \nab )  \bn ({\bf R}) \cdot \bhua ]  [ (
\br_{12} \cdot \nab )  \bn ({\bf R}) \cdot \bhub ] \nonumber \\
&+& \cdots  ,
\end{eqnarray}
with $\int d { \bf s } ^ { \prime } = \iiiint d { \bf R } d \br_{12}  d \bhua d \bhub $ and $ \dot{f} = \partial f ( \bhui \cdot \bn ({\bf R}) ) / \partial ( \bhui \cdot \bn ({\bf R}) ) $.
Let us now consider a twist deformation of the director with a helix axis  along
the $z$-direction of the laboratory frame. Assuming the local nematic
director to describe a perfect helix, we can parametrize the director
field as follows
\beq
\bn ( {\bf R} ) = \cos (q Z) {\bf \hat{x}}+ \sin(q Z) {\bf \hat{y}}  ,
\eeq
with $q=2\pi/p$ the magnitude of the pitch wave vector and $p$ the length of the
cholesteric
pitch. For small wave vectors, $qa \ll 1 $ (with $a$ the typical range of the
pair potential) the trigonometric functions can be expanded to leading order and the spatial dependence of the director approximated as
\beq
\bn ( {\bf R} ) =  {\bf \hat{x}}+ q Z {\bf \hat{y}} + {\cal O }(q^2)  .
\eeq
The  free energy density of the twisted nematic state can then be expressed as
\begin{eqnarray}
\frac{F^{\text{ex}}}{V } &=& \frac{\rho ^2}{2} \iint d \bhua d \bhub 
M_{0}(\bhua, \bhub) f(\bhua) f(\bhub) \nonumber \\
&& - K_t q + \frac{1}{2} K_{2} q^2  ,\label{ftwist}
\end{eqnarray}
where the coefficients pertain to the twist energy and twist elastic contributions, respectively:
\begin{eqnarray}
K_t (\bn \cdot \nab \times \bn) &=& - \frac{\rho ^2}{2} \iint d \bhua d \bhub M_1 (\bhua, \bhub) \nonumber \\
 && \times  u_{2y}  f(\bhua) \dot{f}(\bhub)  
\label{twistpar1}
\end{eqnarray}
\begin{eqnarray}
K_{2} (\bn \cdot \nab \times \bn)^2 &=& - \frac{\rho ^2}{2} \iint d \bhua d \bhub M_2 (\bhua, \bhub) \nonumber \\
&& \times   u_{1y}u_{2y} \dot{f}(\bhua) \dot{f}(\bhub)  . \label{twistpar2}
\end{eqnarray}
The first contribution, (\eq{twistpar1}), is non-vanishing only if the rod interactions are chiral, as we will see later on. It provides a measure for the torque-field exerted by the microscopic chiral interaction which favors the twist distortion.  The second term, \eq{twistpar2}, provides a microscopic
expression for the twist elastic modulus $K_{2}$  of  the Frank
elastic free energy \cite{frank}. The elastic contribution  is a restoring (Hookian) term in the free energy which resists the twist deformation.
The quantities $M_i(\bhua,\bhub)$ are given by the following {\em moment} spatial integrals over the interaction kernel $\Phi$:
\beq
M_{k}(\bhua, \bhub) =  \int d  \br_{12}   \Phi ( \br_{12}  ; \bhua, \bhub) z_{12}^{k}, \hspace{0.3cm} (k=0,1,2)  .\label{moment}
\eeq
In case of achiral hard particles, only the reference term in \eq{ftwist} needs to be considered.
Moreover, with $M_{0}(\bhua,\bhub) = v_{\text{excl}}(\bhua,\bhub)$, the excluded
volume of a pair of particles, the original Onsager free energy \cite{onsager} is recovered as required.

For chiral nematic systems the equilibrium value of the cholesteric pitch is found by
balancing the chiral forces inducing the twist deformation with the elastic forces which favor the nematic. Minimizing the total free energy with respect to the local ODF and $q$ leads to following coupled set of stationarity conditions:
\begin{eqnarray}
\frac{\delta }{\delta f (\bhu) } \left (
\frac{F^{\text{id}}+F^{\text{ex}}}{V} - \mu \int d \bhu f(\bhu) \right ) &=& 0 \label{formal} \\
\frac{\partial }{\partial q } \left ( 
\frac{F^{\text{id}}+F^{\text{ex}}}{V} \right ) &=& 0  ,
\end{eqnarray}
where $\mu $ is a Lagrange multiplier associated with the normalization constraint for the ODF ($\int d \bhu f(\bhu) \equiv 1$) and
\beq
\frac{F^{\text{id}}}{V} = \kbt \int d \bhu \rho f (\bhu) [ \ln {\cal V} \rho f(\bhu) - 1] \label{ideal}
\eeq
is the exact ideal free energy of a spatially uniform system (${\cal V}$
is the thermal volume of the rod).
For weak twist deformations  it is safe to assume that the local ODF remains unaffected by the twist and that it adopts the same form as in the nematic.
 Denoting $f(\bhu)=f_{0}(\bhu)$, the equilibrium ODF of the nematic phase, the equilibrium value for the pitch wave vector at a given density $\rho$ is given by the ratio of the average chiral and elastic forces~\cite{gennes-prost}:
\beq
q  = \frac{ K_{t}[f_{0}]}{K_{2}[f_{0}]}  .
\label{q}
\eeq
The expressions for the twist parameters can be made analytically tractable by using a Gaussian trial function Ansatz to describe the local ODF \cite{OdijkLekkerkerker,odijkchiral}. This is done in Sec. IV. First we have to specify the form of the  interaction potential and kernel $\Phi$ based on a suitable pair potential for chiral spherocylinders.

\section{Generalized van der Waals theory for SW rods}

Whilst the seminal view of Onsager~\cite{onsager} that the repulsive inflexible core of a particle gives rise to orientationally ordered phases is now very well established, the specific nature of the dispersive and polar interactions can have an important influence on the macroscopic structures that are observed. For example, in the case of hard rods with central point dipoles, layered liquid crystalline phases such as the smectic-A are favored, while the effect on the isotropic-nematic transition
appears to be small and in some cases unfavorable \cite{mcgrother1, mcgrother2, mcgrother3}; for molecules
with terminal dipoles, the nematic phase is stabilized relative the smectic phase \cite{mcgrother4}. As can be inferred from the discussion in the introductory section, the influence of attractive interactions is all the more beguiling and subtle in the case of systems with chiral interactions, where the pitch of the helix is found to be very sensitive to the balance of forces that the molecules experience.

Let us  consider an ensemble of hard spherocylinders (HSC), cylinders of length $L$ capped by hemispheres of diameter $D$.  The rods can be rendered chiral by introducing the simple pseudo-scalar potential, proposed by Goossens \cite{goossens}. For any non-overlapping configuration of a pair of rods the chiral contribution reads
\beq
 \Phi_{\text{chiral}} = - \varepsilon_{212} u(  r_{12} ) T_{212} (\hat{\br}_{12} ; \bhua, \bhub ),
\eeq
which consists of a radial part $u(r_{12})$ and an orientation-dependent  pseudo-scalar defined as
\beq
T_{212} ( \hat{\br}_{12} ; \bhua,  \bhub )=  ( \bhua \cdot \bhub) (\bhua \times \bhub \cdot \hat{\br}_{12}), \label{t212}
\eeq
which in fact represents the first non-trivial term in a series expansion in terms of generalized chiral pseudo-scalars $ T_{2i(2k-1)2j}$. For the present case of weak chiral interactions it suffices to retain only the first term.  As $T_{212}$ changes sign  upon interchanging particle positions, $ \hat{\br}_{12} \rightarrow - \hat{\br }_{12} $, while  keeping the orientations fixed,  the pseudo-scalar imparts a chiral interaction. The sign of the amplitude $\varepsilon_{212}$ defines the handedness of the chiral interaction and the corresponding helical mesostructure.

In  Goossens' model \cite{goossens} the radial part decays steeply via $u(r_{12}) = 1/r_{12}^{7}$, which arises from a summation over electrostatic dipolar interaction sites located on each rod. Here, we employ a much simpler dependence based on a simple SW form with range $\lambda$. We thus specify  $u(r_{12})=H(\lambda - r_{12})$, with $H$ a Heaviside step function. To take into account the effect of {\em achiral} attractive (or soft-repulsive) forces we introduce an additional achiral SW potential with amplitude (well depth) $\varepsilon_{000}$. For simplicity, we assume both the chiral and achiral SW potentials to have the same interaction range $\lambda$:
\begin{eqnarray}
 \Phi_{\text{achiral}} &=& - \varepsilon_{000}  H( \lambda  - r_{12}   )
\nonumber \\
 \Phi_{\text{chiral}} &=& - \varepsilon_{212}  T_{212} (\hat{\br}_{12} ; \bhua, \bhub )  H( \lambda  - r_{12}   ),
\end{eqnarray}
for any non-overlapping rod pair configuration.
Putting all contributions together, we arrive at the following total pair potential for  chiral SW rods:
\beq
 \Phi_{\text{tot}} (\br_{12}; \bhua, \bhub) =
  \begin{cases}
\infty & r_{12} < \sigma \\
   - \varepsilon _{000} - \varepsilon_{212} T_{212} & \sigma  \leq r_{12} < \lambda  \\
    0 &  r_{12} \geq \lambda   , \\
  \end{cases}
\eeq
with $\sigma ( \hat{\br}_{12} ; \bhua, \bhub)$ the centre-of-mass  contact distance between two hard spherocylinders at given (relative) orientations. Henceforth, we shall fix the SW range at $\lambda = L + D$ (the so-called ``square peg in a round hole''  model). It is advantageous to introduce a {\em reduced temperature}
defined as
$T^{\ast} = k_{B}T/ | \varepsilon_{000} | $.
An expression for the free energy of the nematic phase of particles consisting of a hard anisotropic core with attractive interactions can be obtained from a first-order perturbation theory  around a suitable hard-core reference free energy. A generalized van der Waals (GvdW) form can be expressed as \cite{gelbartbaron,francomelgarthesis,francomelgar}
\begin{eqnarray}
 \label{gvdw}
\frac{F^{\text{ex}}_{\text{GvdW}}}{V} &=& \frac{ \rho ^2 }{2}\kbt G(\phi) 
\iint d \bhua  d \bhub f(\bhua) f(\bhub) \nonumber \\
&& \times  \int d \hat{\br}_{12}  \int _{0} ^{\sigma } d r _{12}  r_{12}^{2}  \nonumber \\
& - &  \frac{\rho^2}{2} \iint d \bhua d \bhub f(\bhua)  f(\bhub) \nonumber \\
&& \times  \int d \hat{\br}_{12} \int _{\sigma}^{\lambda} d r _{12}  r_{12}^{2} ( \varepsilon _{000} + \varepsilon_{212} T_{212} ). \qquad 
\end{eqnarray}
The first contribution is the Onsager-Parsons excess free energy 
accounting for 
the hard-core repulsive part of the pair potential. It is based upon a scaled second virial approximation according to the Parsons-Lee approach \cite{parsons,Lee87,Lee89}. It involves a mapping of the radial distribution function for the anisotropic particles onto that of an equivalent hard-sphere system via the virial equation. The rod free energy can ultimately be  linked to the Carnahan-Starling~\cite{carnahanstarling,hansenmcdonald} expression for hard spheres which  provides a simple strategy to account for the effect of higher-body interactions, albeit in an implicit and approximate manner.  The scaling factor
\beq
G( \phi) = \frac{1 - \frac{3}{4} \phi } { (1-\phi)^2 },
\eeq
reduces to unity in the Onsager limit $L/D \rightarrow \infty $ where the packing fraction $\phi$ of the nematic phase at the ordering transition  vanishes. The second term of  \eq{gvdw} is the contribution to the free energy due to the attractive interactions, at the mean-field level of description (i.e., any correlations in the particle positions are neglected). This type of augmented van der Waals equation of state is commonly employed in studies of homogeneous fluids (e.g., see references \cite{jackson1986,green}), and can even be used to describe complex fluid phase equilibria in a quantitative manner (for instance the liquid-liquid phase behavior of hydrocarbons and perfluoroalkanes \cite{archer}, or aqueous mixtures of hydrocarbons \cite{galindo} and amphiphiles \cite{garcialisbona}).

\begin{figure*}
\begin{center}
\includegraphics[clip=,width= 0.45\columnwidth, angle = -90 ]{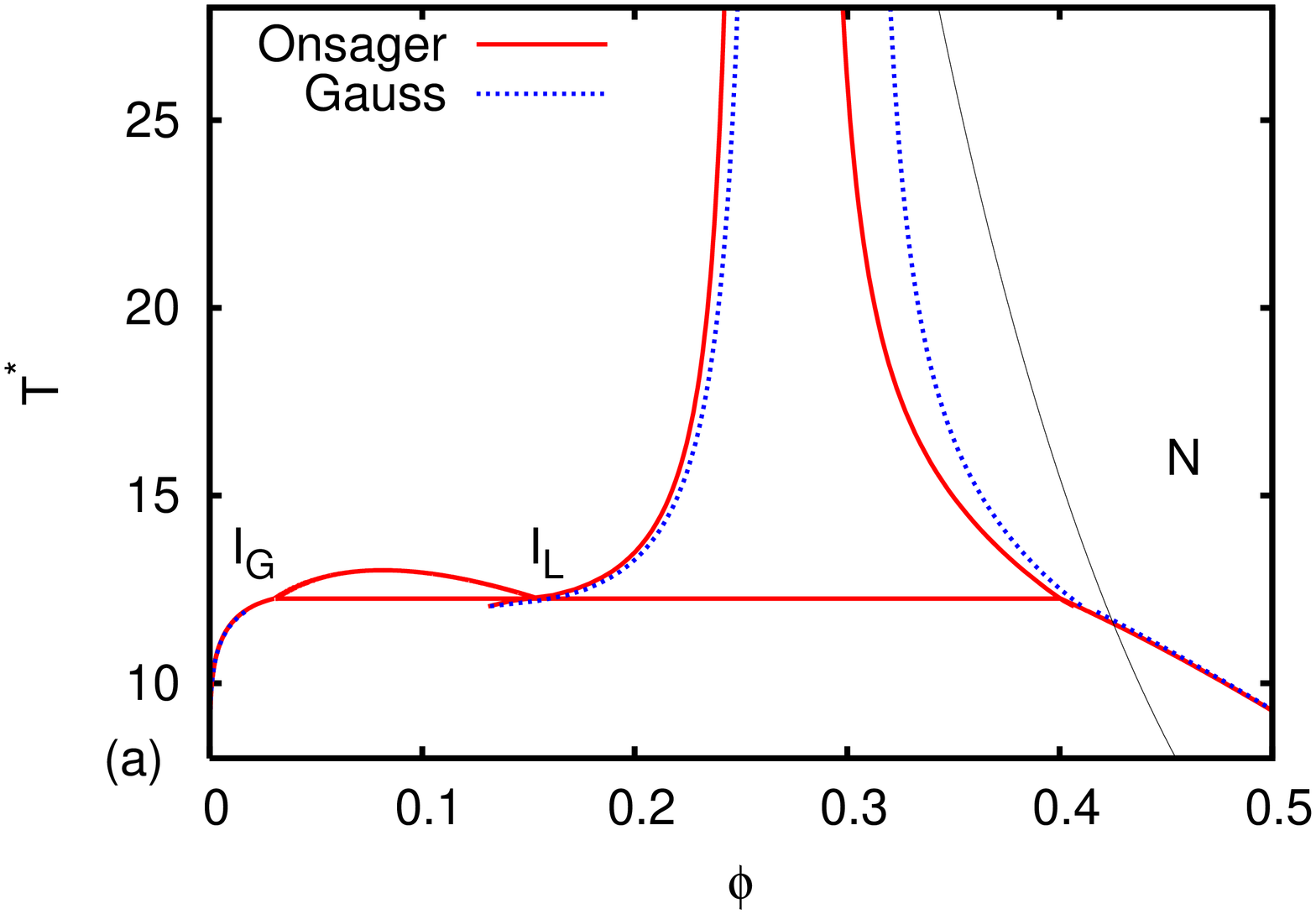}
\includegraphics[clip=,width= 0.45\columnwidth, angle = -90 ]{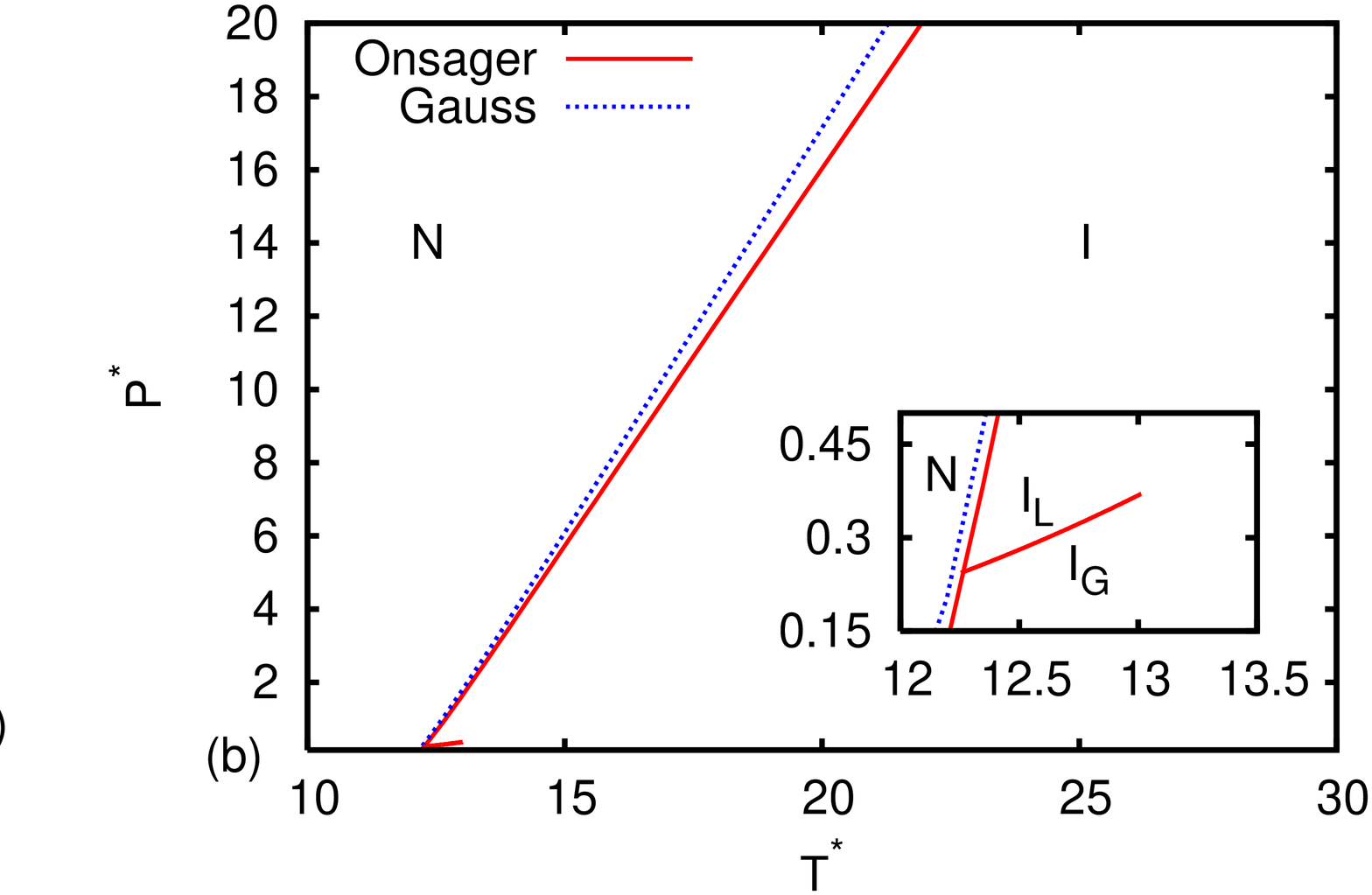}
\includegraphics[clip=,width= 0.45  \columnwidth, angle = -90 ]{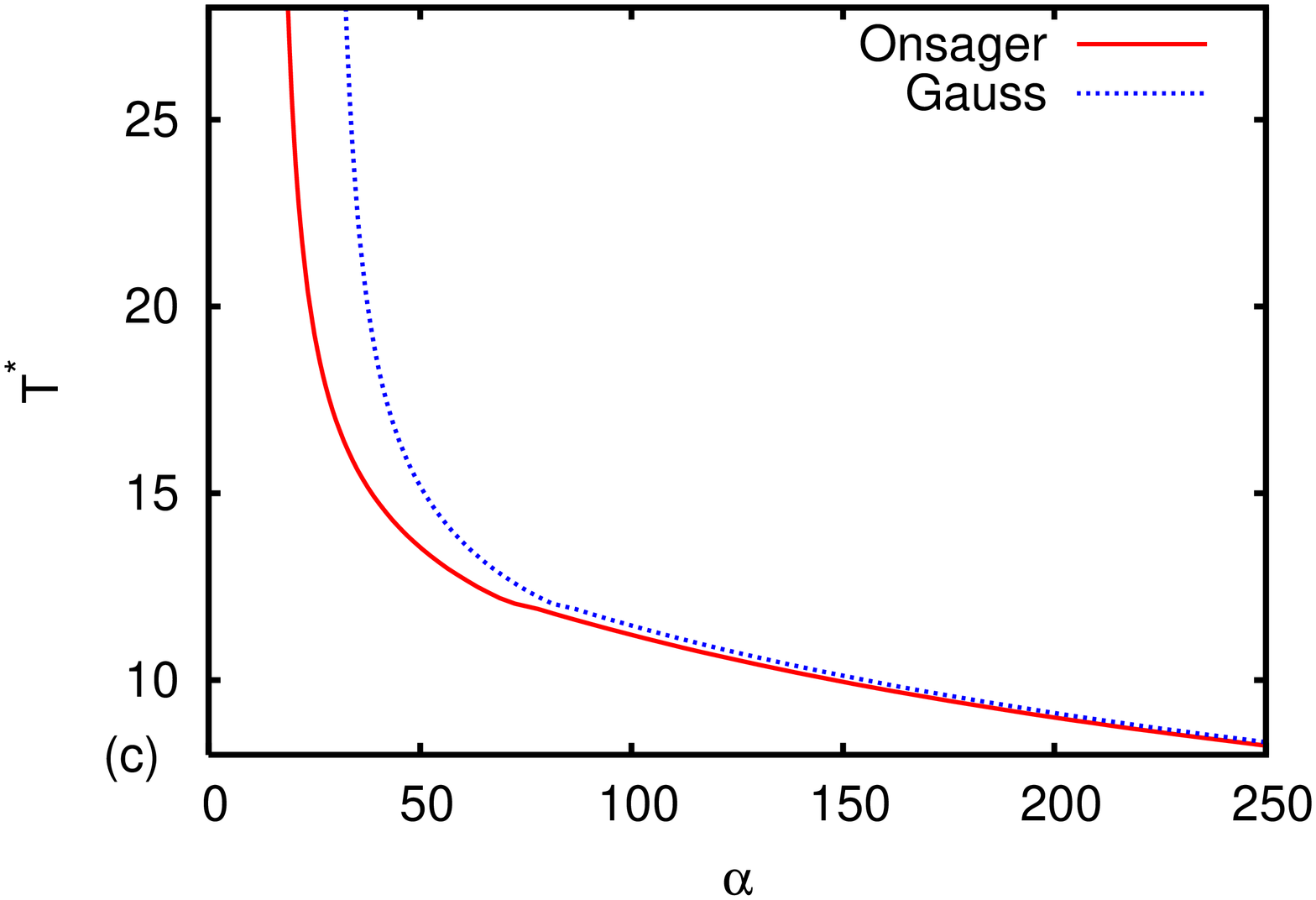}
\caption{ \label{spirh} Phase diagram of achiral ($\varepsilon _{212} = 0$), attractive SW spherocylinders
($\varepsilon _{000} > 0$) of aspect ratio $L/D=10$ and range $\lambda=L+D$. The phase behavior is represented in (a) the reduced temperature $T^{\ast}=k_{B}T/|\varepsilon_{000}|$  versus molecular packing fraction $\phi = v_{0} \rho $ plane with $v_{0}= (\pi/4) LD^2 + (\pi/6)D^3$  the spherocylinder volume, and (b) the reduced pressure $P^{\ast} = Pv_{0}/|\varepsilon_{000}| $ versus $T^{\ast}$ plane; the continuous curve in (a) represents an isobar for $P^{\ast}=120$. The calculations are based on the Onsager and Gaussian trial functions, \eq{ons} and \eq{gauss}, respectively. (c) Variational parameter $\alpha$ of the coexisting nematic phase versus $T^{\ast}$. } 
\end{center}
\end{figure*}

If the GvdW excess free energy is mapped onto the general form given by \eq{fgen} 
we
can obtain the following expression for the total interaction kernel $\Phi $:
\beq
 \Phi ( \br_{12}  ; \bhua, \bhub) = 
  \begin{cases}
  \kbt  G(\phi)   &  r_{12} < \sigma  \\
  - \varepsilon_{000} - \varepsilon_{212}T_{212} &  \sigma \le r_{12} < \lambda. \label{phigvdw}
  \end{cases}
\eeq
It is important to note that  $\Phi $ is not invariant with respect to $ \br_{12} \rightarrow - \br_{12}$ as would be the case for achiral interactions. In the case of ranges of the SW interaction that are at least as long as the longest dimension of the particles ($L+D$) the integration of the interaction decouples neatly into a repulsive and attractive contributions; the integration is not as straightforward for shorter ranged interactions.

The  fluid phase behavior for achiral attractive rods ($\varepsilon _{212} =0 $) has been
discussed extensively  in \olcites{francomelgarthesis,francomelgar} and a typical example is reproduced in \fig{spirh}. For this calculation we only need the zeroth moment $M_{0}$  which  readily follows from \eq{moment} and \eq{phigvdw}:
\beq
M_{0}(\bhua , \bhub) =  (\kbt G(\phi) + \varepsilon_{000} ) v_{\text{excl}}(\gamma) - \varepsilon_{000} \frac{4 \pi}{3} \lambda ^3  ,\label{m0}
\eeq
where $v_{\text{excl}}$ is the excluded volume between two hard
spherocylinders at a relative angle $\gamma = \arcsin |\bhua \times \bhub |$:
\begin{eqnarray}
v_{\text{excl}}(\gamma )  &=&  \frac{1}{3} \int d \hat{\br}_{12} \sigma ^{3}    \nonumber \\
& = & 2 L^{2} D |\sin \gamma | + 2 \pi  D^{2} L + \frac{4 \pi}{3} D^{3}  .
\end{eqnarray}
The ODF $f_{N}(\bhu)$ of the nematic state is formally given by the  solution of the self-consistency equation emerging from \eq{formal}:
\beq
f_{N}( \bhu ) = \frac{\exp \left [- \rho \int d \bhu^{\prime} M_{0}(\bhu ,\bhu^{\prime}) f_{N}(\bhu^{\prime}) \right  ] }{\int d \bhu \exp \left [ -\rho  \int d \bhu^{\prime} M_{0}(\bhu ,\bhu^{\prime}) f_{N}(\bhu^{\prime}) \right  ]}  ,\label{fn}
\eeq
which is not amenable to further analysis and is solved numerically.
The double orientational averages appearing in \eq{gvdw} can be made analytically tractable by adopting a  simple algebraic trial form for the ODF.
In his original paper Onsager \cite{onsager} introduced the following  form to describe the distribution of angles in the uniaxial nematic state:
\beq
f_{O}(\theta) = \frac{\alpha \cosh ( \alpha \cos \theta ) }{4 \pi \sinh \alpha}  , \label{ons}
\eeq
where $ \cos \theta = \bhu \cdot \bn$ is the polar angle between the axis of symmetry of the particle and
the director and  $\alpha \geq  0$ a
variational parameter measuring the degree of nematic order. Note that $\alpha \equiv 0$ in the isotropic phase. The implications of \eq{ons} on the GvdW free energy have been studied in detail in \olcite{francomelgar}.
A simpler trial function has been proposed by Odijk and Lekkerkerker \cite{OdijkLekkerkerker}, based on a Gaussian:
 \beq
f_{G}(\theta) = \frac{\alpha}{4 \pi}
\exp \left ( - \frac{1}{2} \alpha \theta ^2 \right ) \hspace{1cm} 0 \leq \theta \leq \pi/2  ,
\label{gauss}
\eeq
and its mirrored version for the interval $\pi/2 \leq \theta \leq \pi$. Unlike \eq{ons}, the Gaussian trial form does not reduce to the correct isotropic constant $1/4\pi$ for $\alpha = 0$  and we must therefore require $\alpha \gg 1 $ for reasons of consistency. In fact, \eq{ons} becomes identical to the Gaussian form in the asymptotic limit $\alpha \rightarrow \infty$. The results from both distributions will therefore be virtually indistinguishable for large $\alpha$, as is illustrated in \fig{spirh}c.

Let us quote the following Gaussian  averages \cite{OdijkLekkerkerker}:
\begin{eqnarray}
\langle \ln f(\bhu) \rangle &\sim & \ln 4 \pi \alpha - 1  \nonumber \\
\langle \langle \sin \gamma  \rangle \rangle & \sim & \left ( \frac{\pi}{\alpha} \right )^{1/2} \hspace{0.3cm} (\alpha \gg 1), \label{gav}
\end{eqnarray}
where the brackets denote orientational averages: 
\bea
\langle \cdot \rangle &=& \int d \bhu f_{G}(\bhu), \nonumber \\
\langle \langle \cdot \rangle \rangle &=& \iint d \bhua d \bhub f_{G}(\bhua) f_{G}(\bhub).
\eea 
The nematic order parameter $S$ can be approximated by:
\beq
\label{norder}
S \equiv \left \langle  {\cal P}_{2}(\cos \theta) \right \rangle \sim 1 - \frac{3}{\alpha}, 
\eeq 
with ${\cal P}_{2}$ the second-order Legendre polynomial. Using the asymptotic expressions from \eq{gav} in 
the ideal and excess free energy yields a simple algebraic expression for 
the 
free energy. The corresponding phase equilibria can be analyzed without difficulty and the resulting phase diagrams are shown in \fig{spirh}; the coexisting densities are obtained by numerically solving the equality of pressure $P_i=-(\partial F/\partial V)_{N,T}$ and chemical potential $\mu_i=-(\partial F/\partial N)_{V,T}$ of each phase $i$.  At low to moderate temperatures, a coexistence between isotropic gas ($I_{G}$) and liquid  ($I_{L}$) phases is found  in addition to the isotropic-nematic phase separation seen at higher densities. At the triple point temperature $T^{\ast}=12.254$ the isotropic liquid binodal  meets a triphasic $I_{G}$-$I_{L}$-$N$ equilibrium line.
For systems with chiral interactions  ($\varepsilon_{212} \neq 0$) the phase behavior
is
altered by the twist contributions which we shall examine in the next Section.

\section{Calculation of the twist parameters}

The central task in the description of chiral nematic phases within our GvdW theory is the calculation of the moment integrals pertaining to the twist energy and elastic modulus in \eq{twistpar1} and \eq{twistpar2} using the explicit interaction kernel defined in \eq{phigvdw}.  Details of the specific treatment are provided in the next subsections.

\subsection{Twist elastic modulus}

The calculation of the twist elastic modulus consists of two
steps. First, an explicit expression for the second moment
$M_{2}$ in \eq{moment} is required. Then, a double orientational average has to
be carried out using an appropriate form for the ODF according to
\eq{twistpar2}. Inserting \eq{phigvdw} into \eq{moment} and rearranging
terms leads to
†\begin{eqnarray}
M_{2}(\bhua ,\bhub ) & = & (\kbt G(\phi) + \varepsilon_{000}) \int _{v_{\text{excl}}} d \br_{12} z_{12}^{2} \nonumber \\
&&  - \varepsilon_{000} \int _{\lambda} d \br_{12} z_{12}^{2} + {\cal O}(\varepsilon_{212}), \label{m2}
\end{eqnarray}
where $\int _{v_{\text{excl}}} d \br_{12}  = \int d \hat{\br}_{12} \int _{0}^{\sigma} d r_{12} r_{12}^{2} $ denotes a spatial integral over the spherocylinder excluded volume and $\int _{\lambda} d \br_{12}  = \int d \hat{\br}_{12} \int _{0}^{\lambda} d r_{12} r_{12}^{2} $ a spatial integration over the SW range.
All contributions of ${\cal O}(\varepsilon_{212})$ are of negligible importance for weak chirality; more specifically, the ratio of the chiral and non-chiral interactions must
 be small, i.e.,
\beq
\epsilon_{c} =  \left | \frac{ \varepsilon_{212}  }{ \varepsilon_{000} } \right | \ll 1  ,
\eeq
where one should also note that $G(\phi)>1$. The {\em chirality parameter} $\epsilon_{c}$ is,
in principle, fixed by the detailed molecular structure of the rod-like
mesogen and is expected to be small for most common chiral substances. In the case of strong chiral interactions,  [$\epsilon_c \sim {\cal O}(1)$], the twist elastic response will be affected by $T_{212}$ (and higher order pseudo-scalar potentials)  which severely complicates the analysis. The second integral in \eq{m2} is easily evaluated by exploiting the symmetry of the SW potential along the $z$-axis and using cylindrical coordinates:
\beq
\int _{\lambda} d  \br_{12} z_{12}^{2}  =  2 \pi \int_{-\lambda}^{\lambda} d z_{12} z_{12}^{2} \int_{0}^{\sqrt{ \lambda^{2} - z_{12}^{2} }} dr r = \frac{4 \pi \lambda ^{5}}{15} ,
\eeq
where $r=(x_{12}^{2}+y_{12}^{2})^{1/2}$. The first term in \eq{m2} involves a weighted spatial integral over the
excluded-volume manifold spanned by two spherocylinders at fixed
orientations.  Details of this calculation can be found in  Appendix A. The final expression for $M_{2}$ reads:
\begin{widetext}
\begin{eqnarray}
M_{2} (\bhua , \bhub) & = & ( \kbt G(\phi)  + \varepsilon_{000} ) \left [  L^{2} D |\sin \gamma | \right . \frac{2}{3} \left ( A_{1}^{2} + A_{2}^{2} + B^{2} \right ) \nonumber \\
&& +  LD^{2} \left \{ \frac{16}{3} ( A_{1} C_{2} - A_{2} C_{1} ) + \frac{8 \pi}{3} (A_{1}^{2} + A_{2}^{2})
 +  \left .  \pi B^2 + \frac{\pi}{2} ( C_{1}^{2} + C_{2}^{2} ) \right \} + v_{M_{2}}^{HH}  \right ]  - \frac{4 \pi \varepsilon_{000} \lambda ^{5}}{15},\nonumber \\ \label{m2final}
\end{eqnarray}
\end{widetext}
where $v_{M_{2}}^{HH} \propto {\cal O}(D^5)$ is given by \eq{vm2hh} in Appendix A.  The second moment excluded volume is independent of the chiral interaction in the limit of infinitesimally small twist distortions considered here. In \eq{m2final}, $A_{i}$, $B$ and $C_{i}$ are orientationally dependent dot products specified in \eq{dot} of Appendix A.

The elastic modulus is obtained from a double orientational average of $M_{2}$ for which we shall invoke a Gaussian trial function for the ODF, {\em cf.} \eq{gauss}.
In principle a twist deformation of the director field breaks the uniaxial
symmetry of the local ODF, which would now take on a biaxial form, and  a suitable
generalization of the Gaussian distribution involving an explicit dependence on
the azimuthal angle  would therefore be required. To keep
the theory  tractable, we shall ignore local biaxiality and use the original form of \eq{gauss}. As the degree of biaxial nematic order is very small in the weak deformation limit considered here, it is  unlikely to have a significant effect on the mesoscopic properties of the cholesteric phase.

Since the Gaussian ODF is only appropriate if the orientational distribution is strongly peaked around the nematic director  we may perform an asymptotic expansion
of the Gaussian integrals to extract the leading order contributions for
large $\alpha $.  Let us fix $\bz = \{0,0,1\}$ and introduce
\beq
\bhu_{i} = \{ \cos \theta _i, \sin \theta _i \cos \varphi _i , \sin \theta _i \sin \varphi _i  \},
\eeq
the orientational unit vector of rod $i=1,2$ in terms of the polar ($\theta_i$) and azimuthal ($\varphi_i$) angle with respect to the nematic director
which we have fixed along the $x$-direction of the Cartesian frame $\bn =\bx= \{1,0,0\} $. Expanding the dot products in \eq{dot} for $\theta_i \ll 1 $ we obtain up to leading order:
\begin{eqnarray}
|\sin \gamma | & \sim &  |\gamma| \sim  \left ( \theta_{1}^{2} + \theta _{2}^{2} - 2 \theta_{1} \theta _{2} \cos \Delta \varphi \right ) ^{1/2} \label{singamma} \\
A_i & \sim & (L/2) \theta_{i} \sin \varphi_i  \nonumber \\
B & \sim &  D (  \theta _{2} \cos \varphi _{2} - \theta_{1} \cos \varphi_{1} ) / | \gamma |  \nonumber \\
C_{1} = C_{2} & \sim &  D   ( \theta _{1} \sin \varphi _{1} - \theta_{2} \sin \varphi_{2} ) / | \gamma |  , \label{asdot}
\end{eqnarray}
with $\Delta \varphi = \varphi_2 - \varphi_1$.  The twist elastic modulus can be recast into a finite series in terms of the inverse aspect ratio $x = D/L$:
\begin{eqnarray}
\frac{K_{2}D}{\kbt} & \sim & -c^{2} \xi \left \{ H_{0} + H_{1} x  + H_{2}x^{2} + H_{3}x^{3}  + H_{4} x^{4} \right \} \nonumber \\
&&  + c^{2} \varepsilon_{000} \lambda ^{5}  H_{\varepsilon},
\label{series}
\end{eqnarray}
with $c  = \rho L^2 D $ the dimensionless rod concentration and
\beq
\xi = G(\phi) \pm \frac{1}{T^{\ast}}  ,
\eeq
where $(+)$ applies to an attractive square-well ($\varepsilon_{000}>0$) 
and $(-)$ a  soft-repulsive  square-shoulder ($\varepsilon_{000}<0$) potential.
The coefficients $H_{k}$ represent double Gaussian averages involving the following angular  quantities up to leading order in $\alpha \gg 1$:
\begin{eqnarray}
H_{0} &\sim & \frac{\alpha^2}{12} \left \langle \left \langle {\cal G} | \gamma | (\theta _{1} ^{2} \sin ^{2} \varphi_{1} + \theta_{2}^{2} \sin ^{2} \varphi_{2} ) \right \rangle \right \rangle  + \cdots \nonumber \\
H_{1} &\sim & \frac{\pi \alpha^2}{3} \left \langle \left \langle {\cal G} (\theta _{1} ^{2} \sin ^{2} \varphi_{1} + \theta_{2}^{2} \sin ^{2} \varphi_{2} ) \right \rangle \right \rangle + \cdots  \nonumber \\
H_{2} & \sim & \frac{\alpha^2}{3} \left \langle \left \langle {\cal G} (\theta _{2}  \cos \varphi_{2} - \theta_{1} \cos \varphi_{1} )^{2}/ | \gamma | \right \rangle \right \rangle \nonumber \\
      && + \frac{4 \alpha^2}{3} \left \langle \left \langle {\cal G} (\theta _{1}  \sin \varphi_{1} - \theta_{2} \sin \varphi_{2} )^{2} / | \gamma | \right \rangle \right \rangle  + \cdots \nonumber \\
H_{3} & \sim & \frac{\pi \alpha^2}{2} \left \langle \left \langle {\cal G} ( \theta _{2}  \cos \varphi_{2} - \theta_{1} \cos \varphi_{1} )^{2} / \gamma ^{2} \right \rangle \right \rangle \nonumber \\
      && + \frac{\pi \alpha^2}{2} \left \langle \left \langle {\cal G} ( \theta _{1}  \sin \varphi_{1} - \theta_{2} \sin \varphi_{2} )^{2} / \gamma ^{2} \right \rangle \right \rangle + \cdots \nonumber \\
 H_{4} & \sim & \frac{2\pi \alpha^{2}}{15} \left \langle \left \langle {\cal G} \right \rangle \right \rangle + \cdots   \nonumber \\
H_{\varepsilon} & \sim & \frac{ 2 \pi \alpha^2}{15} \left \langle \left
\langle { \cal G } \right \rangle \right \rangle = 0, \label{haka}
\end{eqnarray}
where 
\beq
{\cal G} \sim  \theta_{1} \theta_{2} \cos \varphi_{1} \cos \varphi_{2} + \cdots 
\eeq
The brackets represent the following four-fold  angular integral:
\beq
 \left \langle \left \langle \cdot \right \rangle \right \rangle \sim \alpha^2 \prod_{i=1,2} \int_{0}^{\infty} d \theta_{i} \theta_{i} \exp \left [ - \frac{ \alpha }{2} \theta _{i}^{2} \right ]  \int _{0}^{2 \pi}
 \left ( \frac{d  \varphi_i}{2 \pi} \right ), \label{bracket}
\eeq
where we have used  $\dot{f}_{G} = \partial f_{G} / \partial (\cos \theta ) \sim \alpha  f_{G} $. It is evident that $H_{\varepsilon}$ vanishes upon integration over the azimuthal angles, irrespective of the form of the (uniaxial) ODF. A little inspection shows that the leading order  contributions to $H_{1}$ and $H_{4}$ are also eliminated by the double azimuthal integration. The remaining terms are, in principle, nonzero because $| \gamma | $  depends non-randomly on $\Delta \varphi $.  The results can be greatly simplified by changing to the new azimuthal variables
$\varphi_{1} = \psi $
and $\varphi_{2}  = \psi + \Delta \varphi $. The integration over $\psi$ can be carried out without difficulty. After some algebra the averages reduce to:
\begin{eqnarray}
H_{0} & \sim & \frac{\alpha^2}{96} \left \langle \left \langle \theta_{1} \theta_{2} (\theta _{1}^{2} + \theta_{2}^{2} ) | \gamma | \cos \Delta \varphi \right \rangle \right \rangle  + \cdots \nonumber \\
H_{2} & \sim & \frac{\alpha^2}{24} \left \langle \left \langle  \theta_{1}
\theta_{2} \left [ (\theta_{1}^{2} + \theta_{2}^{2} ) 7 \cos \Delta \varphi  \right . \right . \right . \nonumber \\
&& \left .  \left . \left .  - (4 + 10 \cos 2 \Delta \varphi) \right ] /|\gamma |  \right \rangle \right \rangle + \cdots \nonumber \\
H_{3} & \sim & \frac{\pi \alpha^2}{4}\left \langle \left \langle \theta_{1} \theta_{2} \cos \Delta \varphi \right \rangle \right \rangle + \cdots \label{hs}
\end{eqnarray}
Here, the brackets now denote \eq{bracket} with the double azimuthal integral over $\varphi_{1,2}$ replaced by a single one over the remaining angle $\Delta \varphi$. Clearly, the azimuthal average yields $H_{3}=0$  so that we need only evaluate the two  even contributions $H_{0}$ and $H_{2}$. The first term corresponding to infinitely long rods has been analyzed by Odijk in \olcite{odijkelastic}. Eliminating $\cos \Delta \varphi$ via the asymptotic expression for $| \sin \gamma| $ [\eq{singamma}] leads to: 
\beq
H_{0} \sim  \frac{\alpha^2}{96} \left [ - \left \langle \left \langle | \gamma | ^{3} \theta_{1}^{2} \right \rangle \right \rangle  + \left \langle \left \langle | \gamma | \theta_{1}^{2} (\theta_{1}^{2} + \theta_{2}^{2} )   \right \rangle \right \rangle  \right ]   \label{h0e}
\eeq 
We may now use the Gaussian averages in Appendix B to arrive at the compact expression,
\beq
H_{0}   \sim  -\frac{7}{192 } \left ( \frac{\pi}{\alpha} \right ) ^{1/2} 
\eeq
Similarly, by applying \eq{singamma} to $H_{2}$ we may simplify: 
\bea
H_{2} & \sim & \frac{\alpha^2}{24} \left [ \frac{13}{24} \left \langle \left \langle | \gamma | \theta_{1}^{2}  \right \rangle \right \rangle   
- \frac{5}{24} \left \langle \left \langle | \gamma | ^{3}  \right \rangle \right \rangle \right . \nonumber \\
&& \left . + \frac{1}{8} \left \langle \left \langle  \theta_{1}^{2}  (\theta_{2}^{2} - \theta_{1}^{2}) /  |\gamma|  \right \rangle \right \rangle  \right ] \label{h2e}.
\eea
The first two Gaussian averages are given in Appendix B. The last one cannot be calculated analytically, but the $\alpha$ dependence is easily established from a simple analysis of the scaling, while the pre-factor can be found numerically (see Appendix B). The final result is:
\beq
H_{2} \sim  -\kappa \alpha ^{1/2},
\eeq
with $ \kappa=0.036926 $.
For later reference, we also give the next leading order $\alpha$-contributions to $H_{k}$.
It can be shown that the odd terms $H_1$ and $H_3$ in \eq{haka} are fully eliminated (to all order in $\alpha$) by the double azimuthal integration. 
The higher order $\alpha$-contributions to $H_{k}$ (even $k$) are in principle non-vanishing and can be estimated by considering the angular quantities in \eq{haka} taking the next power in the polar angle $\theta_{i}$.  From simple scaling considerations it then follows that the corrections $\delta H_{k}$ must be at least of order:
\bea
\delta H_{0} & \sim & {\cal O}(\alpha^{-1}), \nonumber \\
\delta H_{2} & \sim & {\cal O}(\alpha^{-1/2}), \nonumber \\
\delta H_{4} &\sim & {\cal O}(\alpha^{1/2}).\label{nlo}
\eea
which all give marginal contributions to $K_{2}$ for large aspect ratio. 

The equilibrium value for $\alpha$ is obtained from
the reference free energy of the (undistorted) nematic. Combining the 
ideal and excess parts of the nematic free energy,  \eq{ideal} and \eq{ftwist} respectively, with the expression for $M_{0}$ [\eq{m0}] gives: 
\beq
\frac{F^{\text{GvdW}}}{\kbt N} = \left \langle  \ln {\cal V} \rho f
(\bhu)  - 1 \right \rangle  + \frac{\rho \xi}{2} \left \langle \left \langle v_{\text{excl}} ( \bhua, \bhub ) \right \rangle \right \rangle  +  \frac{2 \pi  \rho \lambda ^{3}}{3 T^{\ast}}.
\eeq
On inserting the Gaussian averages for the ideal and excluded volume contributions [\eq{gav}],  the asymptotic free energy becomes
\beq
\frac{F^{\text{GvdW}}}{\kbt N} \sim \ln \alpha  + c \xi (\pi / \alpha )^{1/2},
\eeq
where all terms independent of $\alpha$ have been omitted for compactness as they do not contribute to the degree of orientational order. Minimizing the free energy with respect to $\alpha$ yields the common quadratic form \cite{Vroege92}:
\beq
\label{ac2}
\alpha  \sim (\pi^{1/2} c\xi /2)^{2} .
\eeq
With this relation for $\alpha$,  an  {\em algebraic} expression for the twist elastic constant for the SW spherocylinders can be formulated from \eq{series}. After defining  the spherocylinder packing fraction as $\phi \simeq (\pi/4)xc$ (ignoring the end-cap corrections), we obtain
\beq
K_2^{\ast}=\frac{K_{2}D}{\kbt} x  \sim \phi \left \{ \frac{7}{24
\pi}  + \frac{32}{\pi^{5/2}} \kappa (\phi \xi)^{2} + {\cal O}(x) \right \} \label{k2} .
\eeq
The twist elastic modulus primarily depends  on the particle packing fraction
and the reduced temperature, with the aspect ratio merely playing the role of a linear scaling factor.

\begin{figure}
\begin{center}
\includegraphics[clip=,width= 0.7 \columnwidth, angle = -90 ]{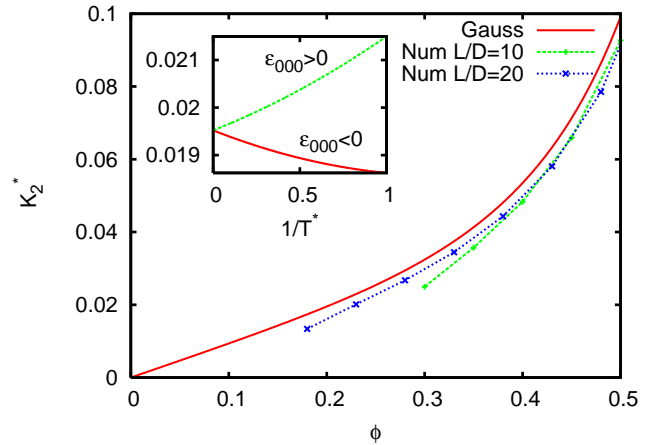}
\caption{ \label{twist} Twist elastic modulus $ K_2^{\ast}=K_{2}Dx / \kbt $ [\eq{k2}] for hard spherocylinders ($1/T^{\ast} =0$) as a function of packing fraction $\phi = v_0 \rho$. The numerical ODF is given by the solution of \eq{fn}. Inset: Temperature dependence of $ K_{2}^{\ast} $ for rods of aspect ratio $L/D=20$ with attractive square-well ($\varepsilon_{000} > 0$) and soft-repulsive square-shoulder ($\varepsilon_{000}<0$) interactions of range $\lambda=L+D$ at a fixed packing fraction of $\phi = 0.2$. }
\end{center}
\end{figure}

Some remarks are now in order. The leading order contribution was found 
for infinitely thin rods \cite{odijkelastic} and does not depend on 
temperature. The twist elastic modulus is {\em independent} of the range 
$\lambda$ of the SW potential. Correction terms arising from the next 
leading order terms in \eq{nlo} are at least of order $x=D/L$  and thus of marginal influence for sufficiently slender rods. 
\fig{twist} shows that the discrepancy with numerical results is very small even for relatively short rods with $L/D=10$.  
The numerical data are based on 
a numerical evaluation of \eq{twistpar2} using the exact ODF from 
\eq{fn}. It is important to note that the cholesteric phase is only stable with 
respect to the isotropic state roughly when $c \gtrsim 1 $ or equivalently 
$\phi \gtrsim x $. As illustrated in \fig{twist}, numerical 
results for the nematic solution of the ODF \eq{fn} are found only above a 
critical packing fraction.

\begin{figure*}
\begin{center}
\includegraphics[clip=,width= 0.7\columnwidth, angle = -90 ]{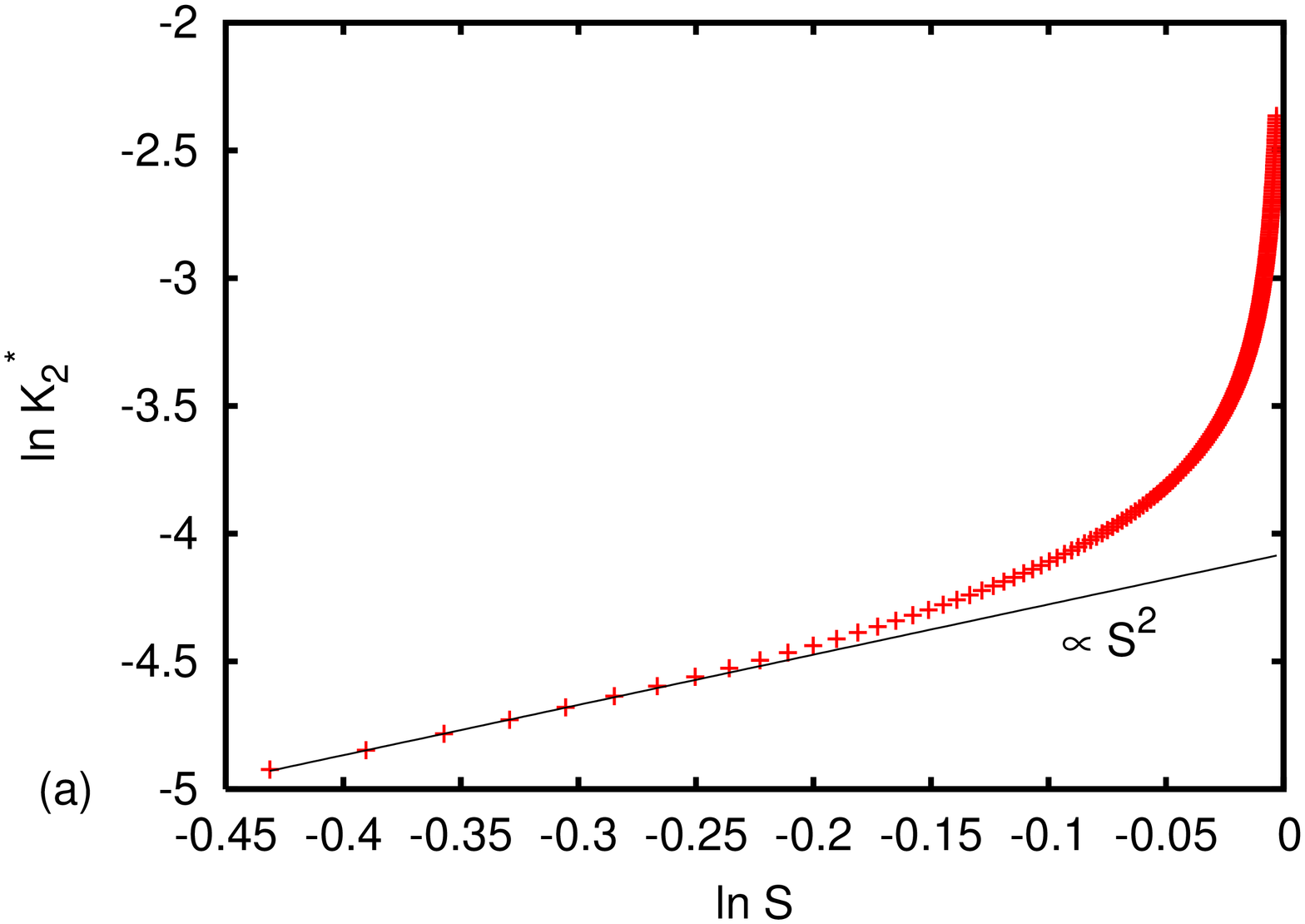}
\includegraphics[clip=,width= 0.7\columnwidth, angle = -90 ]{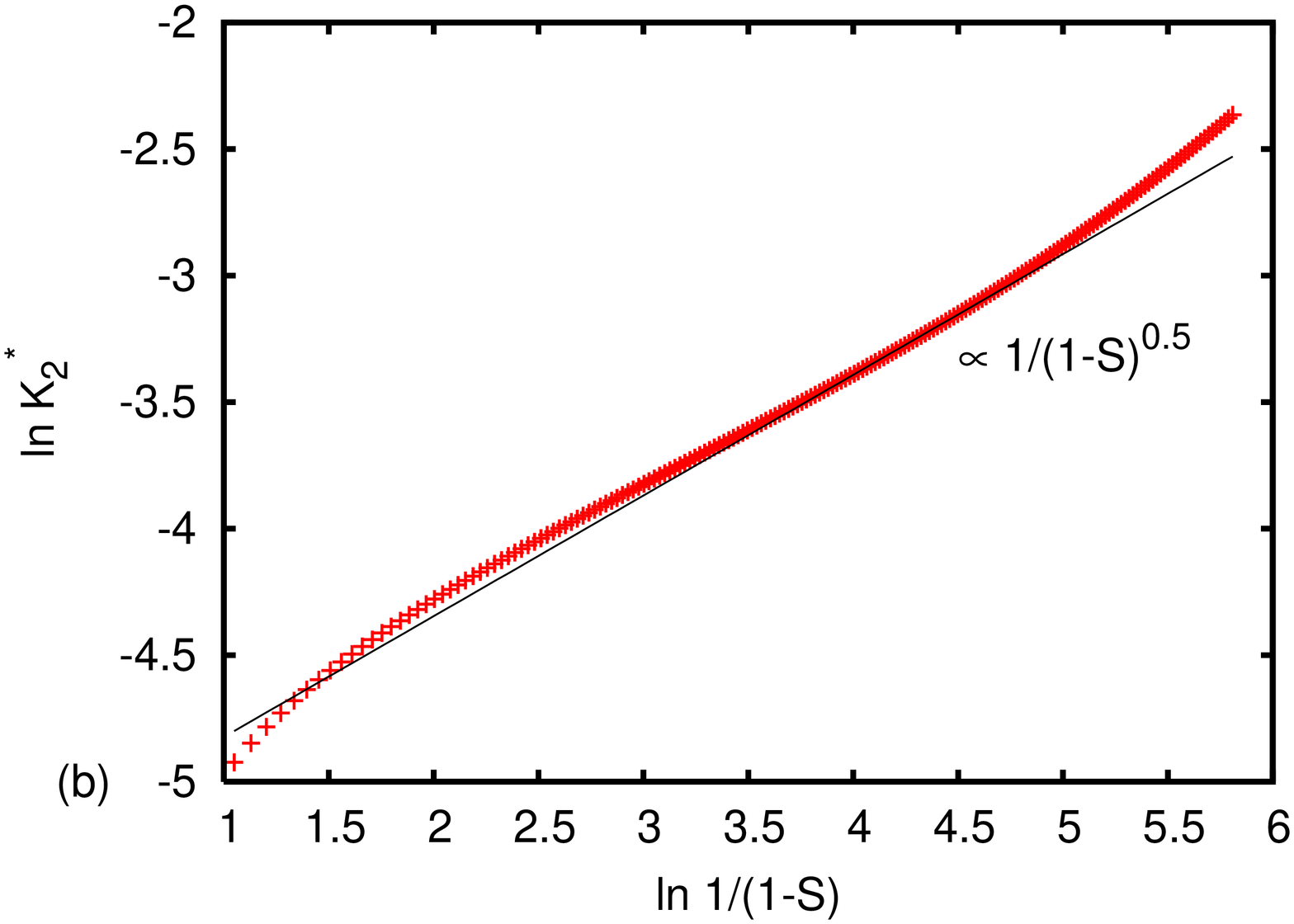}
\caption{ \label{twistss}  Scaling of the twist elastic modulus $K_{2}^{\ast}$ with respect to the nematic order parameter $S$. The results are obtained from the numerical ODF [\eq{fn}] using $L/D=20$. Continuous lines represent linear fits. (a) Quadratic behavior $K_{2}^{\ast} \sim S^{2} $ at moderate nematic order $0.6<S<0.9$. (b) Asymptotic behavior $K_{2}^{\ast} \sim  1/(1-S) ^{\omega}$ with $\omega \ge 0.5$  in the regime of high nematic order ($S>0.99$). }
\end{center}
\end{figure*}

As expected, the twist elastic modulus is a monotonically increasing 
function of the packing fraction. The temperature dependence shown in 
\fig{twist} indicates an increase of the twist elastic resistance for 
attractive rods ($\varepsilon_{000} > 0$), while the opposite trend is 
observed for a soft-repulsive square-shoulder potential 
($\varepsilon_{000} < 0$). The latter interaction may be particularly 
suitable for the description of colloidal $fd$-virus rods as a crude model 
for the electric double layer or the polymer coat grafted onto the colloid 
surface \cite{grelet-fraden_chol}.

The behavior of the twist elastic modulus with respect to the nematic order parameter is highlighted in \fig{twistss}.
Since $\alpha \sim \phi^{2}$ [\eq{ac2}] and $\alpha \sim 1/(1-S)$ [\eq{norder}] it is readily deduced that the asymptotic behavior of the twist elastic modulus for strong nematic order is given by the following scaling relation:
\beq
\label{scaless}
K_{2}^{\ast} \sim a \left  ( \frac{1}{1-S} \right )^{1/2} + b \left (\frac{1}{1-S} \right  ) ^{3/2},
\eeq
in terms of the constants $a$ and $b$. This scaling relation is confirmed in \fig{twistss}b based on numerical results for the twist elastic modulus \cite{lee-elastic}. The crossover between the two scaling contributions in \eq{scaless} is reflected by the fact that the exponent [$1/(1-S)^{\omega}$] lies within the range $0.5 < \omega < 1.5$ as $S$ approaches unity (or equivalently $1/(1-S) \rightarrow \infty $). A different scaling behavior is observed for moderate nematic order where   $K_{2}^{\ast}$ is found to be proportional to $S^2$ (\fig{twistss}a). This result is in agreement with previous theoretical predictions based on a Legendre series expansion of the angular properties valid for weak nematic order \cite{priest,gelbart}.

\subsection{Twist energy}

We proceed in a similar manner for the calculation of the twist energy. The first-moment spatial integral over the interaction kernel \eq{phigvdw} is required first. None of the terms pertaining to the achiral parts of the interaction potential contribute to the twist energy so that
\begin{eqnarray}
M_{1}(\bhua ,\bhub ) &=&   \varepsilon_{212} \int _{v_{\text{excl}}} d \br_{12} z_{12} T _{212}  \nonumber \\
   && -  \varepsilon_{212} \int _{\lambda} d  \br_{12} z_{12} T_{212},   \label{m1}
\end{eqnarray}
The second spatial integral  is easily tackled by switching to cylindrical coordinates. Let us write $\br_{12} = r \sin \beta {\bf \hat{x}} +  r \cos \beta {\bf \hat{y}} +  z_{12} {\bf \hat{z}}$ so that
\begin{eqnarray}
    \int _{\lambda} d \br_{12} z_{12} T_{212} &=& \int _{0}^{2 \pi} d \beta \int_{-\lambda}^{\lambda} d z_{12} z_{12} \nonumber \\ 
&& \times \int_{0}^{\sqrt{ \lambda^{2} - z_{12}^{2} }} dr r  T_{212} \nonumber \\
&=& \frac{\pi}{3} \lambda ^{4} (\bhua \cdot \bhub) (\bhua \times \bhub \cdot {\bf \hat{z}})  .
\end{eqnarray}
The first  spatial integral runs over the excluded volume of the spherocylinder for which we use the parametrization advanced in Appendix A.
This produces terms of  ${\cal O}(LD^{3})$ and higher order in $D$ which we will not show explicitly. We can thus write the first-moment integral as the following expression:
\beq
M_{1}(\bhua ,\bhub ) = -\frac{\pi}{3} \varepsilon_{212} \lambda ^{4} (\bhua \cdot \bhub) (\bhua \times \bhub \cdot {\bf \hat{z}}) + {\cal O}(LD^{3}) .
\eeq
Inserting the asymptotic forms of $\bhu_{i}$ for small polar angles into \eq{twistpar1} and performing the azimuthal integration over $\varphi_{1}$  one can express the chiral torque as
\begin{eqnarray}
\frac{K_{t}D^{2}}{\kbt}  & \sim &    \frac{\pi}{6} c^{2} \varepsilon_{212} \left ( \frac{\lambda}{L} \right )^{4} \alpha \left  \langle  - \theta_{2}^{2} \cos^{2} \varphi_{2} \right \rangle  + {\cal O}(x^{3}) \nonumber \\
& \sim & \frac{\pi}{6} \frac{\epsilon_{c}}{T^{\ast}} c^{2} \left (  \frac{\lambda}{L} \right )^{4} + {\cal O}(x^{3}) \label{kt}  ,
\end{eqnarray}
where the elementary  Gaussian average $\langle \theta ^{2} \cos^{2} \varphi \rangle \sim 1/\alpha $ has been employed. This expression is similar to the one derived for infinitely thin rods \cite{odijkelastic}. The correction terms due to finite rod thickness are deemed to be very small, and can be ignored if the spherocylinders are not too short.

\subsection{Cholesteric pitch}

A microscopic expression for the cholesteric pitch of SW spherocylinders in the limit of weak twist distortion is  obtained by combining \eq{k2} and \eq{kt} in \eq{q}. After some rearranging one can write the pitch $p=2\pi/q$ in a convenient reduced form as
\bea
\label{pitch}
p^{\ast} & = & \left ( \frac{p}{L} \right ) \bar{\epsilon} \nonumber \\
 & \sim & \frac{T^{\ast}}{\phi} \left \{  \frac{7 \pi}{32} + \frac{24 \kappa \phi^{2}}{\pi^{1/2}} \left ( G(\phi) \pm \frac{1}{T^{\ast}} \right )^{2} + {\cal O}(x) \right \}, \qquad
\eea
where $(+)$ refers to a square-well and $(-)$ to a square-shoulder potential. The factor $\bar{\epsilon}$ scaling the pitch combines  the chirality parameter $\epsilon_{c}$ and the geometric parameters of the range $\lambda$ and (inverse) aspect ratio $x=D/L \ll 1$. It can be interpreted as an integrated van der Waals energy which depends on the fourth power of the (SW) interaction range $ \lambda $:
\beq
\bar{\epsilon} = \frac{ \epsilon_{c} } {x^{2}} \left ( \frac{\lambda}{L}\right )^{4}.
\eeq
As with the twist elastic modulus,  \eq{k2}, the rescaled pitch  depends only on the packing fraction and reduced temperature of the cholesteric phase.

\section{Results and discussion}

\begin{figure*}
\begin{center}
\includegraphics[clip=,width= 0.7\columnwidth, angle = -90 ]{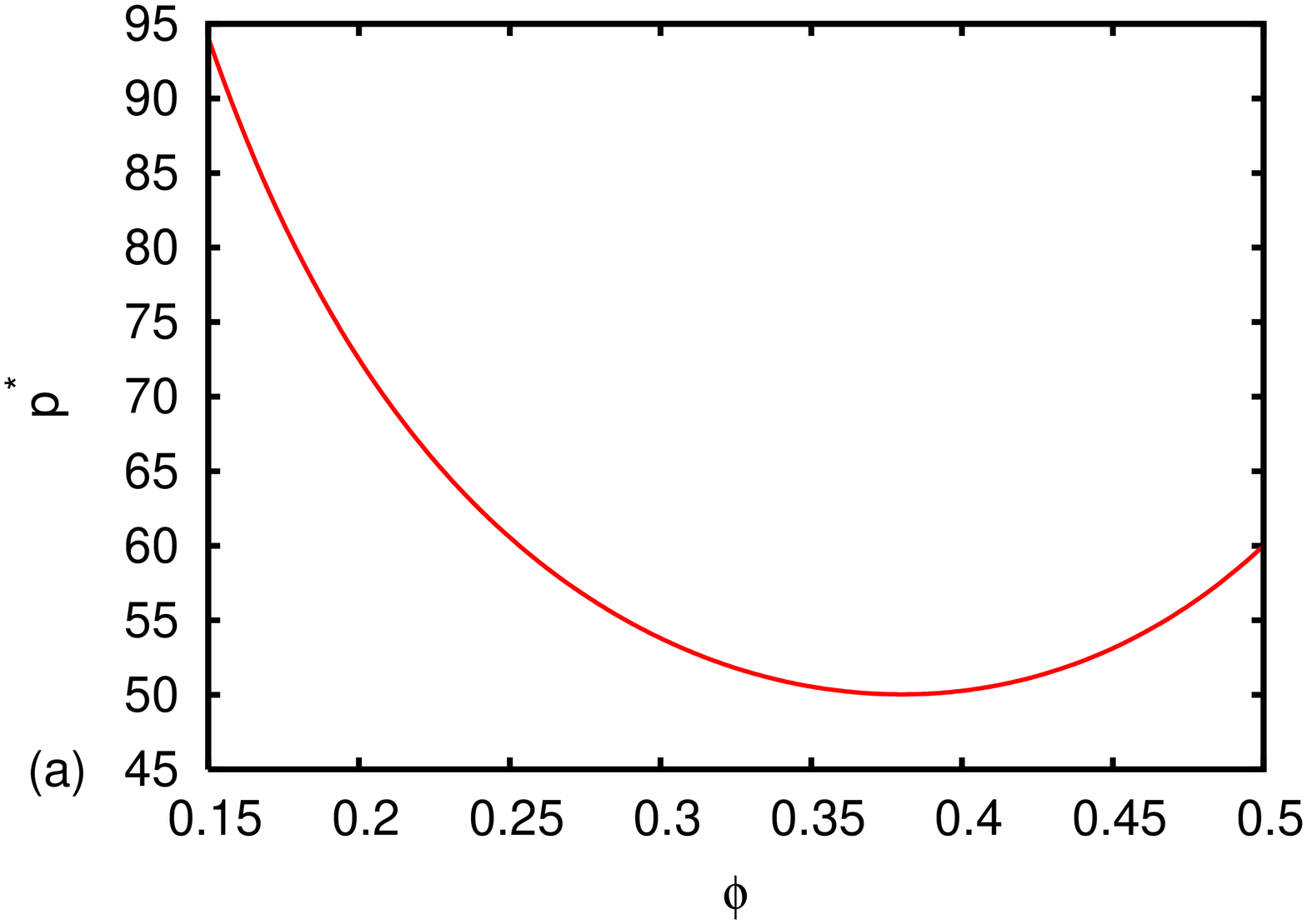}
\includegraphics[clip=,width= 0.7\columnwidth, angle = -90 ]{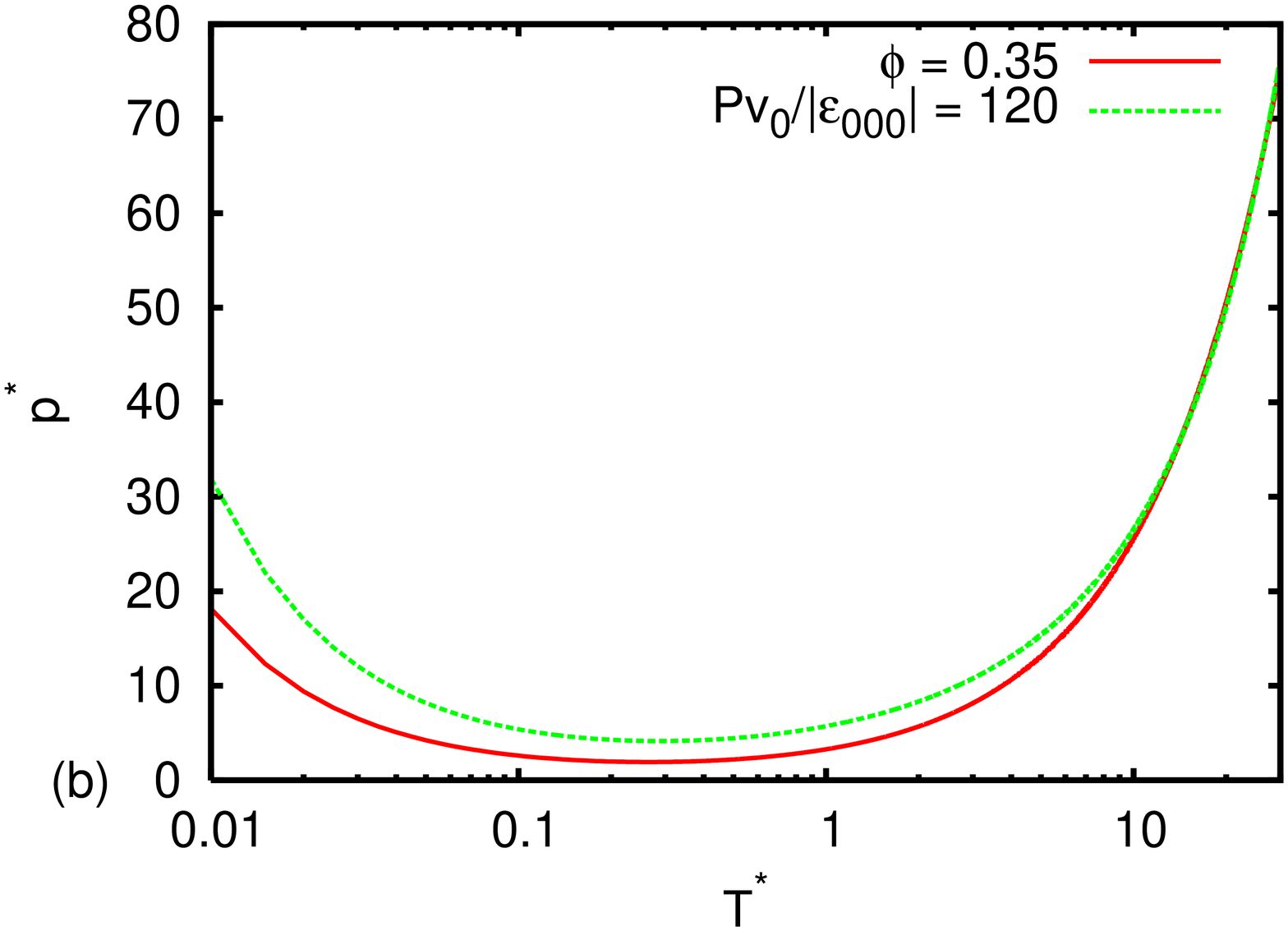}
\caption{ \label{pitchfig} (a) Scaled cholesteric pitch $ p^{\ast}= (p/L) 
\bar{\epsilon}$   [\eq{pitch}] for attractive SW spherocylinders at 
reduced temperature  $T^{\ast} =\kbt/|\varepsilon_{000}|=20$  versus rod 
packing fraction $ \phi = v_0 \rho $. (b) Variation of the scaled pitch as 
a function of $T^{\ast}$ (on a logscale) for a constant packing fraction 
$\phi=0.35$. 
Also shown is the behavior for a fixed rescaled pressure $P^{\ast}=Pv_{0}/|\varepsilon_{000}|=120$ corresponding to the isobar (thin continuous curve) in \fig{spirh}a. }\end{center}
\end{figure*}

The density and temperature dependence of the cholesteric pitch obtained with the theory presented in the previous sections is depicted in \fig{pitchfig}. The variation of the pitch with packing fraction is non-monotonic, irrespective of the reduced temperature. The steep decrease at low densities is a common result for slender rods. Since $K_{t} \propto \phi ^2 $ and $K_{2} \propto \phi $ [{\em cf.} \eq{kt} and \eq{k2}], the free-energy cost associated with the elastic deformation upon incrementing the density is more than offset by the simultaneous free energy gain due to an increase of the torque-field associated with the chiral interactions.  The  inverse proportionality $p \propto \phi^{-1} $ is in agreement with Odijk's result for rigid rods \cite{odijkchiral} and consistent with experimental results of $fd$-rods at high ionic strength \cite{grelet-fraden_chol}. At higher packing fractions, the orientational (nematic) order rises sharply and the influence of the finite rod thickness, embodied by the second term in \eq{pitch}, becomes apparent. The elastic resistance rises steeply and causes the cholesteric structure to unwind upon increasing density.

For large temperatures ($1/T^{\ast}<1$),  the behavior of the pitch is only weakly affected by the nature of the non-chiral interactions as can be inferred from the last term in \eq{pitch}.  From a qualitative point of view, the behavior is therefore the same for attractive square-well and soft repulsive square-shoulder potentials. Moreover, since $p^{\ast} \propto T^{\ast}$, the temperature merely serves as a linear scaling factor and does not influence the shape of the curve in \fig{pitchfig}a.

\begin{figure*}
\begin{center}
\includegraphics[clip=,width= 0.7\columnwidth, angle = -90 ]{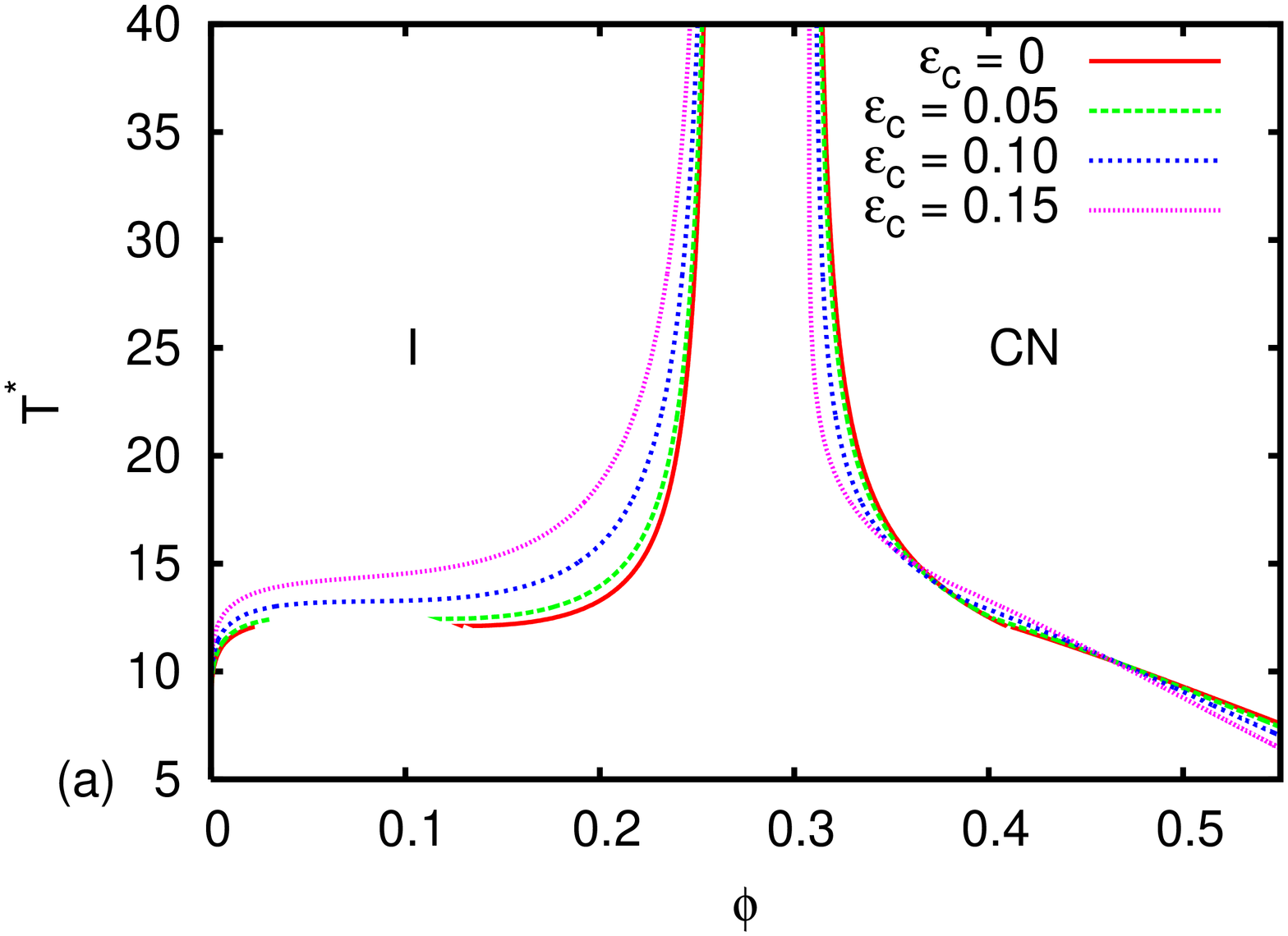}
\includegraphics[clip=,width= 0.7\columnwidth, angle = -90 ]{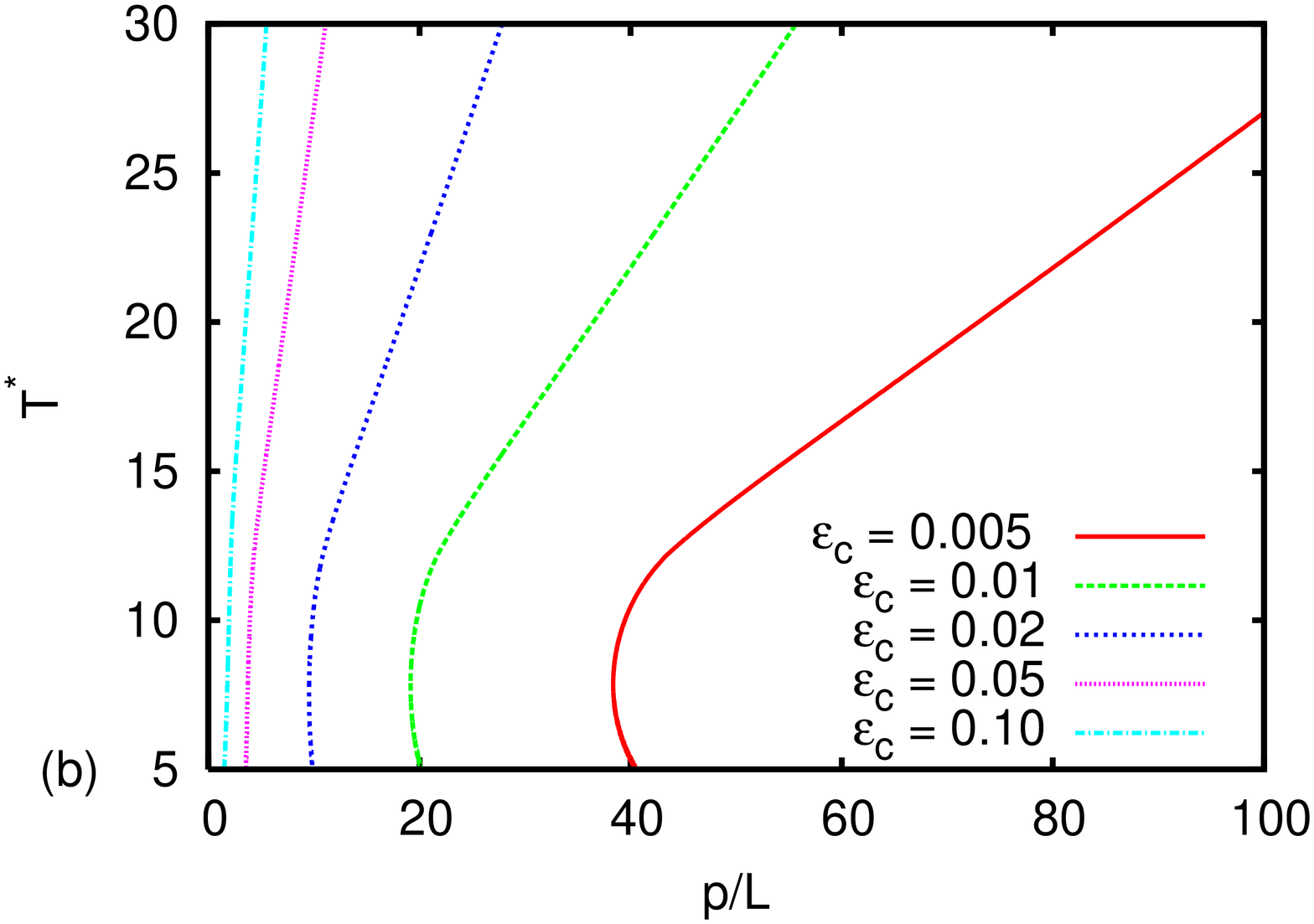}
\caption{ \label{fdchiral} (a) Phase diagram for attractive chiral SW spherocylinders ($L/D=10$) of range $\lambda=L+D$  for various relative strengths of the chiral interaction $\epsilon_{c}= | \varepsilon_{202}/\varepsilon_{000} |$, where $T^{\ast}=\kbt/|\varepsilon_{000}|$  is the reduced temperature and $\phi = v_0 \rho$ the packing fraction. $CN$ denotes the cholesteric phase. (b) Temperature dependence of the reduced pitch $p^{\ast}=p/L$ of the cholesteric phase in coexistence with the isotropic. }
\end{center}
\end{figure*}

It has been suggested that the increase of the pitch with density is due to pre-smectic fluctuations that counteract the twist deformation \cite{ alben,pindak,gennes-prost}. The present analysis would suggest that the observation  can be accounted for within a simple mean-field theory for anisotropic yet homogeneous systems in which one disregards density fluctuations; layered structures such as the smectic-A (SmA) phase possess inhomogeneities in the average particle position. That  smectic fluctuations are not necessary to give rise to an increase in the pitch is also supported by experimental observations in colloidal $fd$-rods \cite{dogic-fraden_chol} where an unwinding of the cholesteric is observed at densities far below the cholesteric-smectic transition density.
 We stress that parallel configurations (induced by the finite rod thickness, i.e., spherocylindrical shape) that lead to an unwinding of the cholesteric state with increasing density also facilitate the formation of a smectic-A phase. The two phenomena are therefore expected to be correlated. The possibility of a transition towards a smectic phase is not incorporated in our theory but it is anticipated that such an instability would lead to a much steeper increase and possibly a divergence of the pitch with density. Based on the phase diagram of pure hard spherocylinders  \cite{mcgrother,Bolhuisintracing} the cholesteric-smectic transition would occur at a packing fraction of  $\phi \sim 0.4 $, irrespective of the aspect ratio (for sufficiently long rods). The unwinding of the cholesteric phase at high densities is not unique to the $fd$ system but has also observed in solutions of polysaccharide and polypeptide compounds \cite{yoshiba-sato} and DNA \cite{strey-dna,rill-dna}.
Theoretically, such a trend  was first established in simulations based on the chiral hard spherocylinders \cite{vargachiral}, supplemented by a simple ground-state theory for a chiral hard Gaussian overlap model assuming perfect local nematic order \cite{vargachiral1}.

The temperature dependence of the pitch is also highly non-monotonic. The near linear increase of the pitch with temperature for a system at constant density (or pressure) is a direct result of the achiral hard-core repulsion  between the rods which  dominates  the (chiral) attractive interactions  at moderate to high temperatures. At $T \rightarrow \infty$ the free energy of the system is governed entirely by the entropic contribution associated with hard spherocylinders, resulting in a nematic phase ($p \rightarrow \infty$).  The unwinding of the pitch with temperature has been found in solutions of polypeptides~\cite{dupreduke75} , and most notably for aqueous suspensions of $fd$-virus rods \cite{dogic-fraden_chol} which was not accompanied by a thermal change in the intrinsic chirality of the viral structure (e.g., due to a denaturation of the protein coating).  Theoretically, this would translate to a chirality parameter $\epsilon_c$ which is constant, as we have assumed here. Our predictions of an increase in pitch with increasing temperature are also consistent with the findings for the cholesterol ester CEEC~\cite{harada}, a thermotropic mesogen which does not exhibit a smectic phase.

The marked increase in pitch at low temperatures is in accordance with experimental observations in numerous thermotropic systems the most ubiquitous being the derivatives of cholesterol  \cite{alben,pindak}. A similar trend albeit less prominent  is found for aqueous solutions of $fd$ virus \cite{dogic-fraden_chol}.   The underlying scenario is analogous to the unwinding of the cholesteric with increasing density for the lyotropic case and originates from a stark increase in nematic order upon lowering $T^{\ast}$. The twist elastic resistance associated with the near-parallel configurations becomes anomalously large and gives rise to a strong unwinding of the cholesteric structure. As for the behavior with density, an additional coupling and a divergent pitch is expected at the nematic-smectic transition temperature as observed by Pindak \etal \cite{pindak} and theoretically advanced by de Gennes \cite{degennes-elastic}. We do point out that  in our particular model the decrease of the pitch with temperature (and constant density or pressure) observed at low temperatures is metastable with respect to an isotropic (gas)-cholesteric phase separation (see the phase diagram in \fig{spirh}a).

 Finally, from \eq{pitch}  one can obtain a simple relation between the pitch and the particle aspect ratio, $p/L \propto x^2 $, or equivalently $p \propto 1/L$. The pitch of the cholesteric thus becomes tighter upon increasing the aspect ratio  at a fixed mass density and temperature. This result is in line with observations in $fd$  systems where the pitch is found to decrease for larger viral contour lengths  \cite{grelet-fraden_chol}. The relation  found in experiment, $p \propto 1/L^{0.25}$,  reveals a much weaker dependency which could be attributed to the slight degree of flexibility, neglected in the present rigid-rod model. It is noteworthy that the Straley-Odijk theory \cite{straleychiral,odijkchiral} predicts the opposite trend $p \propto L$. Their model is based on short-ranged (steric) chiral interactions induced by a thin helical thread of thickness $D$ enveloping the rod.  Here, the chiral interactions are long-ranged  and scale with the rod length $\lambda \propto L$. A linear increase in the pitch with aspect ratio is predicted with a scaled Onsager theory for chiral hard Gaussian overlap rods with the approximation of perfect local nematic order~\cite{vargachiral1}. Although  no microscopic justification for our scaling relation can be given, it does give the correct scaling of the pitch with contour length and is consistent with the long-ranged nature of the chiral forces between $fd$ virus rods \cite{grelet-fraden_chol}.

In \fig{fdchiral} we shows that there is noticeable effect on the phase behavior of attractive SW spherocylinders for weakly chiral interactions. Upon increasing the strength of the chiral interaction relative to the dispersion forces ($ \epsilon_c $) the isotropic-cholesteric transition  shifts to lower densities with an additional broadening of the two-phase coexistence region
at moderate to high temperatures.  Beyond a certain critical value,  the isotropic gas-liquid envelope and the corresponding triple point become metastable relative to the direct coexistence between a low-density isotropic and a high density cholesteric phase. Since the chiral interactions are essentially attractive, we observe the same stabilization of the cholesteric state for repulsive square-shoulder potentials. The non-monotonic behavior of the pitch with temperature is reflected in \fig{fdchiral} where the evolution of the pitch of the coexisting cholesteric phase is depicted. It is clear that pitches ranging from a few molecular lengths to a few hundreds of molecular lengths can be reproduced with our theory.

The quantitative merits of the present theory can only be assessed by comparison with experiment for specific values of the interaction parameters of our model. The most important one is the chirality parameter $\epsilon_c$ which captures the intrinsic strength of the chiral interactions. Apart from the shape of the mesogen, it is determined primarily by the nature of the long-range electrostatic interactions and the intricate surface structure of the particles; in the case of viruses, for example, the chiral interaction will depend on the helical configuration of surface charges. A change in temperature may induce  conformational changes in the chiral structure which affect the magnitude or sign of the pitch. For example, a pitch inversion  involving a sudden change of handedness of the cholesteric structure  with  temperature is known to occur in thermotropic systems \cite{uematsu,sikora}. These issues are clearly beyond the scope of our coarse-grained model and require a much more detailed representation of the molecular architecture of the  mesogen \cite{tombolatoferrarini,kornyshevleikin-chiral}.

\section{Conclusions} An asymptotic analysis of the
deformation free energy associated with a cholesteric phase of chiral
spherocylinders is presented in this work. The rods consist of a hard
spherocylindrical backbone with additional long-ranged achiral
 (attractive/soft repulsive) and chiral interactions, both represented by a
simple square-well form with a range comparable to the length of the rod.
Analytical expressions for the twist elastic constant and cholesteric
pitch are deduced by invoking a Gaussian approximation for the
orientational distribution around the local nematic director.  The
approach is expected to provide an accurate representation of weakly twisted cholesteric states with a high degree
of local nematic ordering which is essentially unaffected by the weak spatial variation of the
director field.

The results are relevant to both thermotropic mesogens (e.g., derivatives of cholesterol) and more particularly lyotropic cholesteric systems such as $fd$
virus rods where chiral interactions are mediated predominantly by
long-range electrostatic forces \cite{grelet-fraden_chol}. The theory
captures the behavior of the cholesteric pitch of $fd$ rods, in particular
its non-monotonic variation with temperature and density, as well as the
influence of the viral contour length (the  effective  aspect ratio). An extension of the theory to simple hard-sphere chain models (e.g., see \olcite{williamson1998}) would enable one to examine the effect of flexibility in more detail.

For attractive SW rods, a steep increase of the pitch is found upon
lowering the temperature, in line with experimental observations in thermotropic systems.  The steep unwinding of
the pitch at low temperatures or high packing fractions is primarily due
to a sharp increase of the local nematic order. The prevailing near-parallel rod
configurations  lead to an anomalous increase of the twist elastic resistance. This simple mean-field scenario contrasts with the commonly expounded  view in which the unwinding of the cholesteric is attributed to pre-smectic fluctuations which are geometrically incompatible with a helical structure.

In the future, we plan to validate our theoretical findings with a simulation study of the current chiral spherocylinder model
along the lines of \olcites{allenmasters,tjiptoelastic,vargachiral}. This will allow us to test
the accuracy of the Onsager-Parsons theory in predicting derivative properties such
as the elastic constants of dense nematic systems. The simulations would also provide a better insight into the behavior of the cholesteric pitch close to a cholesteric-smectic transition. Finally, it would be intriguing to study the implications of the chiral interactions on the micro-structure of the smectic phase.

\section*{Appendix A: Parametrization of the excluded volume of the spherocylinder}

\begin{figure}
\begin{center}
\includegraphics[clip=,width= 0.7 \columnwidth ]{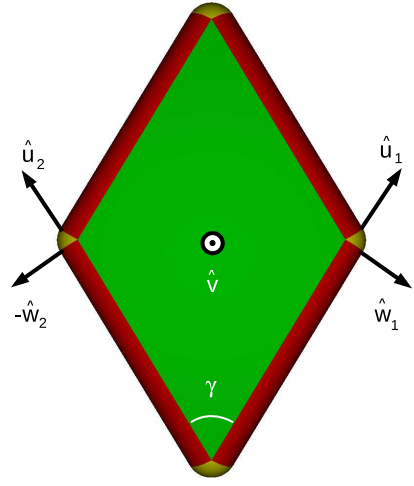}
\caption{ \label{prism} (Color online). Illustration of the excluded volume manifold  of two spherocylinders for a relative orientation angle $\gamma$ and the unit vectors defined in \eq{unit}. The cylinder-cylinder $CC$, cylinder-hemisphere $CH$, and hemisphere-hemisphere $HH$ volumetric sections are indicated in green, red and yellow, respectively. } \end{center}
\end{figure}

The excluded volume manifold of two hard spherocylinders at fixed angle $\gamma$ is a {\em spheroparallelepiped} (see \fig{prism}) which is most conveniently parametrized by switching from the laboratory frame to a particle frame based on the orientational unit vectors $\bhu_{i}$. Let us further define the unit vectors
\begin{eqnarray}
\bv  & = &  \frac{\bhua \times \bhub }{| \sin \gamma |} \nonumber \\
\bw _{i} & = &   \bhu _ {i} \times \bv, \hspace{0.5cm} (i=1,2) \label{unit}
\end{eqnarray}
so that $ \{ \bhu_{i} , \bv , \bw_{i} \}$  are two orthonormal basis sets in 3D. The centre-of-mass distance vector can be uniquely decomposed in terms of these basis vectors:
\beq
\br _{12} =  ( \br _{12} \cdot \bhu_{i} )  \bhu_{i}  +  ( \br _{12} \cdot \bv )  \bv  +  ( \br _{12} \cdot \bw_{i} )  \bw_{i}. \hspace{0.4cm} (i=1,2)
\eeq
The leading order contribution to the excluded-volume body is of
${\cal O}(L^{2}D)$ and stems from the overlap of the cylindrical parts of the spherocylinders. This gives rise to a 3D parallelepiped (central section in \fig{prism}) which can be parametrized as
\beq
\br _{12}^{CC} =  \frac{L}{2} t_1 \bhua  + \frac{L}{2} t_2 \bhub   + D t_3 \bv  ,
\eeq
with  $-1 \le  t_i \le 1 $ for $i=1,2,3$. The Jacobian associated with the coordinate transformation is $J_{CC}= \frac{1}{4} L^{2} D | \sin \gamma | $.
Higher order contributions account for end-cap effects due to the finite
thickness of the rods \cite{vroijthesis}. The first correction term is of ${\cal O}(LD^2)$
and originates from an overlap of the hemispherical end-cap of one rod with the cylindrical part of the other. The resulting  half cylinders on the four edges of the parallelepiped, at the boundaries of the excluded volume in \fig{prism}, can be parametrized as follows:
\begin{eqnarray}
\br _{12}^{CH \pm} & = & \mp \frac{L}{2} \bhub + \frac{L}{2} t_{1} \bhua  + t_{2}D (\pm \bw_{1} \cos \beta  + \bv \sin \beta  ) \nonumber \\
\br _{12}^{HC \pm} & = & \pm \frac{L}{2} \bhua + \frac{L}{2} t_{1} \bhub  + t_{2}D (\pm \bw_{2} \cos \beta  + \bv \sin \beta  ), \nonumber \\
\end{eqnarray}
with boundaries $ -1 \le t_{1} \le 1 $, $ 0 \le t_{2} \le 1$ and $\pi/2
\le \beta \le
\pi/2$ and Jacobian $J_{CH(HC) \pm} = \frac{1}{2} LD^{2}t_{2}$. The final
contribution stems from the  spherical segments located at
the four corners of the parallelepiped (yellow sections in \fig{prism}). These arise from the overlap of two
hemispherical end-caps and are of ${\cal O}(D^3)$. The segments can be
parametrized by invoking a different orthonormal set $\{\bhu_{+},\bhu_{-}, \bv\}$ where
\beq
\bu_{\pm} =  \bhua \pm \bhub , \hspace{0.3cm} \bhu_{\pm} = \frac{\bu_{\pm}}{|\bu_{\pm}|}  ,
\eeq
which leads to the following expression for the centre-of-mass distance corresponding to the four segments:
\begin{eqnarray}
\br _{12}^{HH + \pm} & = & \pm \frac{L}{2}  \bu_{+} + rD (\cos \theta \bv + \sin \theta \sin \varphi_{+} \bhu_{-} \nonumber \\
 && \pm   \sin \theta \cos \varphi_{-} \bhu_{+}  ) \nonumber \\
\br _{12}^{HH - \pm} & = & \pm \frac{L}{2}  \bu_{-} + rD (\cos \theta \bv + \sin \theta \sin \varphi_{-} \bhu_{+}  \nonumber \\
&& \pm   \sin \theta \cos \varphi_{+} \bhu_{-}  ),
\end{eqnarray}
with parameter intervals $0 \le r \le 1$, $0 \le \theta \le \pi$, $( \gamma  - \pi)/2 \le \varphi_{+} \le (\pi-\gamma)/2$ and $-\gamma/2 \le \varphi_{-} \le \gamma/2$. The Jacobian associated with the coordinate transformation is $J_{(HH+\pm)(HH-\pm)} = D^3 r^{2} \sin \theta $.

For the twist elastic modulus we require the second-moment excluded volume [{\em cf.} \eq{m2}]:
\beq
v_{M_{2}}(\bhua, \bhub) =  \int _{v_{\text{excl}}} d \br_{12}  ( {\bf \hat{z}}  \cdot \br_{12} )^{2} \label{vm2}  ,
\eeq
which can be evaluated separately for each  of the excluded volume sections  using the parametrization above. The integrations over the parametrization variables can be worked out without difficulty.  Let us define the following dot products
\begin{eqnarray}
A_i &=& (L/2) \bhu_{i} \cdot \hat{{\bf z}}  \nonumber \\
B &=& D \bv  \cdot \hat{{\bf z}} \nonumber \\
C_{i} &=& D \bw_{i} \cdot \hat{{\bf z}} \nonumber \\
E_{\pm} &=& D  \bhu _{\pm} \cdot \hat{{ \bf z  }}.  \label{dot}
\end{eqnarray}
The cylinder-cylinder $CC$ contribution then reads
\begin{eqnarray}
v_{M_{2}}^{CC} &=& \frac{1}{4} L^{2} D | \sin \gamma | \int_{-1}^{1} dt_{1} \int_{-1}^{1} dt_{2}  \int_{-1}^{1} dt_{3} ( {\bf \hat{z}}  \cdot \br_{CC} )^{2}  \nonumber \\
& = & \frac{2}{3}  L^{2} D | \sin \gamma | (A_{1}^{2} + A_{2}^{2}  + B^2 )  .
\end{eqnarray}
For the cylinder-hemisphere $CH$ contributions we obtain:
\begin{eqnarray}
v_{M_{2}}^{CH} &=& \frac{1}{2} L D^{2}  \int_{-1}^{1} dt_{1} \int_{0}^{1} t_{2} dt_{2}  \int_{-\pi/2}^{\pi/2} d\beta \nonumber \\
&& \times  \left \{  ( {\bf \hat{z}}  \cdot \br_{CH_{+}} )^{2}  + ( {\bf \hat{z}}  \cdot \br_{CH_{-}} )^{2} + \cdots \right \}  \nonumber \\
& = & LD^{2} \left \{ \frac{16}{3} ( A_{1} C_{2} - A_{2} C_{1} ) + \frac{8 \pi}{3} (A_{1}^{2} + A_{2}^{2})  \right . \nonumber \\
&& +  \left . \pi B^2 + \frac{\pi}{2} ( C_{1}^{2} + C_{2}^{2} ) \right \}  .
\end{eqnarray}
Finally, for the hemisphere-hemisphere $HH$ contributions we have:
\begin{eqnarray}
v_{M_{2}}^{HH} &=& D^{3}  \int_{0}^{1} dr r^{2}  \int_{0}^{\pi} d \theta \sin \theta  \nonumber \\
&& \times  \left [   \int_{(\gamma -\pi)/2}^{(\pi - \gamma)/2} d \varphi_{+ }   \left \{  ( {\bf \hat{z}}  \cdot \br_{HH_{++}} )^{2}  + \cdots \right \} \right . \nonumber \\
&& +  \left .  \int _{-\gamma/2}^{\gamma/2} d \varphi_{-}   \left \{  ( {\bf \hat{z}}  \cdot \br_{HH_{-+}} )^{2}  + \cdots \right \}   \right ]  \nonumber \\
& = & D^{3} \left \{  \frac{4}{3} \gamma (E_{-} |\bu_{-}|)^{2}  + \frac{4}{3}(\pi - \gamma)  (E_{+}|\bu_{+}|)^{2} \right . \nonumber \\
&& +  \frac{4 \pi }{15} (E_{+}^{2} + E_{-}^{2} + B^{2})  \nonumber \\
&& + \left .\pi \left ( | \bu _{+}| E_{+}^{2}  \cos \frac{\gamma}{2}  +  |\bu_{-} | E_{-}^{2} \sin \frac{\gamma}{2} \right ) \right \} . \label{vm2hh}
\end{eqnarray}
It is easily verified that all contributions are symmetric under inversion $\bhu_{i} \rightarrow -\bhu_{i}$ and interchanging  $\bhua \leftrightarrow \bhub $, as required.
The total expression \eq{vm2} is obtained by adding  the contributions from the  different sections,  $v_{M_{2}} = v_{M_{2}}^{CC}+ v_{M_{2}}^{CH} + v_{M_{2}}^{HH}$, which provides an analytic result for the moment excluded volume of two spherocylinders of arbitrary aspect ratio.

For the twist energy one requires the first-moment excluded volume over the pseudo-scalar $T_{212}$. Recalling \eq{t212} we get
\beq
v_{M_{1}}(\bhua, \bhub) =  \cos \gamma \int _{v_{\text{excl}}} d \br_{12} ( {\bf \hat{z}}  \cdot \br_{12} ) ( \bhua \times \bhub \cdot \hat{\br}_{12})  \label{vm1}  ,
\eeq
with $\cos \gamma = \bhua \cdot \bhub $. Exploiting the orthogonality of the unit vectors and collecting terms gives for  the leading order contribution:
\beq
v_{M_{1}}^{CC} = \frac{1}{4} L^{2}D B \sin ^{2} \gamma \cos \gamma  {\cal F}_{CC} (x, \gamma )  .
 \eeq
Here,  ${\cal F}_{CC}$ represents a triple integral:
\beq
{\cal F}_{CC}(x, \gamma ) = \prod_{i=1}^{3} \int_{-1}^{1} dt_{i} \frac{t_{3}^{2}}{\sqrt{ \frac{t_{1}^{2}}{4x^{2}} + \frac{t_{2}^{2}}{4x^{2}}+ t_{3}^{2}+ \frac{2 t_{1}t_{2}}{4x^2} \cos \gamma}}  ,
\eeq
which cannot be solved in closed form. In the asymptotic limit  ($\cos \gamma \sim 1$) ${\cal F}_{CC} $ becomes a function of the inverse aspect ratio $x=D/L$ only, and no longer plays a  role in the subsequent angular averaging. It is enlightening to expand the argument as a Taylor series:
\beq
 \frac{t_{3}^{2}}{\sqrt{ \frac{t_{1}^{2}}{4x^{2}} + \frac{t_{2}^{2}}{4x^{2}}+ t_{3}^{2}+ \frac{2 t_{1}t_{2}}{4x^2}}}  = \frac{2t_{3}^{2}}{|t_{1}+t_{2}|}x + {\cal O}(x^{3})  ,
\eeq
from which one finds that the leading order $CC$ contribution is of ${\cal O}(LD^{3})$ and thus of marginal importance for the relevant range of aspect ratios $L/D > 10$.
The integration over the $CH$ parts requires more effort, but the result can be cast in a similar form:
\beq
v_{M_{1}}^{CH} = \frac{1}{2} LD^{2} B | \sin \gamma | \cos \gamma  {\cal F}_{CH} (x, \gamma )  ,
\eeq
with
\beq
{\cal F}_{CH} (x, \gamma ) = \int_{-1}^{1} d t_{1} \int _{0}^{1} d t_{2} t_{2} \int_{-\pi/2}^{\pi/2}  d \beta \sum_{\pm} \frac{2 t_{2}^{2}\sin^{2} \beta }{r_{12}^{CH\pm}}  .
\eeq
Here, $r_{12}^{CH\pm}$ represents the vector norms:
\beq
r_{12}^{CH\pm} = \sqrt{ \frac{1}{4x^2} + \frac{t_{1}^{2}}{4x^2} + t_{2}^{2} \pm \frac{t_{1}}{2x^2} + \frac{t_{2}}{2x} \cos \beta |\sin \gamma |}  .
\eeq
Without further analyzing ${\cal F}_{CH}$ it is evident that the $CH$ contributions are small and at least of ${\cal O}(LD^{3})$. The $HH$ integrations will produce terms of even higher order in $D$ and therefore  the analysis need not be pursued any further.

\section*{Appendix B: Gaussian averages}
The Gaussian averages required for the evaluation of $H_{0}$  and $H_{2}$ [\eq{h0e} and \eq{h2e}] have been deduced by Odijk \cite{odijkelastic}. We quote them here:
\begin{eqnarray}
\left \langle \left \langle  |\gamma | \theta _{1}^{2} ( \theta_{1}^{2} + \theta_{2}^{2} )  \right \rangle \right \rangle & \sim & \frac{35}{2}  \pi^{1/2} \alpha ^{-5/2} \nonumber \\
\left \langle \left \langle  |\gamma |^{3} \theta _{1}^{2}  \right \rangle \right \rangle &\sim & 21  \pi ^{1/2} \alpha ^{-5/2}  \nonumber \\
 \left \langle \left \langle  |\gamma |^{3}   \right \rangle \right \rangle & \sim & 6 \pi^{1/2} \alpha ^{-3/2}  \nonumber \\
  \left \langle \left \langle  |\gamma | \theta_{1}^{2}   \right \rangle \right \rangle & \sim & \frac{5}{2}\pi^{1/2} \alpha ^{-3/2} .
\end{eqnarray}
The last term in \eq{h2e} can be evaluated in two steps. Since $ \theta_{1}^{2} (\theta_{2}^{2} - \theta_{1}^{2})/|\gamma| \sim {\cal O}(\theta ^3)$ a simple Gaussian integral suffices to establish the following scaling result:
\beq
\left \langle \left \langle   \frac { \theta_{1}^{2} (\theta_{2}^{2} - \theta_{1}^{2}) } { |\gamma| } \right \rangle  \right \rangle   \propto  \alpha \int _{0}^{\infty} d \theta \theta ^{4} \exp \left [ - \frac{ \alpha \theta^{2}}{2} \right ] \sim  p \alpha ^{-3/2}  .
\eeq
The pre-factor $p$ is obtained by numerical evaluation of the following triple integral:
\begin{eqnarray}
p &=& \alpha^{7/2} \int _{0} ^{\pi/2}  d \theta _{1}  \theta_{1} \int_{0} ^{\pi/2} d \theta_{2} \theta_{2}  \int _{0}^{2\pi}  \frac{ d \Delta \varphi } {2 \pi} \nonumber \\
&& \times \exp \left [ - \frac{ \alpha }{2} (\theta_{1}^{2} + \theta _{1}^{2} )  \right ] \frac{  \theta_{1}^{2} (\theta_{2}^{2} - \theta_{1}^{2}) } { |\gamma| } ,
\end{eqnarray}
which gives $ p  =-1.772$ for any given $\alpha \gg 1$.

\acknowledgments
We are grateful to Szabolcs Varga for fruitful discussions. HHW acknowledges the Ramsay Memorial Fellowship Trust for financial support. Funding to the Molecular Systems Engineering group from the Engineering and Physical Sciences Research Council (EPSRC) of the UK (grants GR/N35991 and EP/E016340), the Joint Research Equipment Initiative (JREI)
(GR/M94427), and the Royal Society-Wolfson Foundation refurbishment grant is gratefully acknowledged.

\end{document}